%% file: main.tex
\documentclass[twocolumn,tighten]{aastex7} %,linenumbers]

\graphicspath{{./}{}}
\DeclareUnicodeCharacter{2212}{-}
\usepackage[dvipsnames]{xcolor}
\usepackage{multirow}
\usepackage{csvsimple,longtable,booktabs}
\usepackage{enumitem}% http://ctan.org/pkg/enumitem
\usepackage{amsmath}
\begin{document}

\newcommand\editt[1]{{\color{Black}#1}}
\newcommand\edit[1]{{\color{Black}#1}}
\newcommand\response[1]{{\color{Blue}#1}}

\submitjournal{ApJL}
\title{Masses, Star-Formation Efficiencies, and Dynamical Evolution of 18,000 \ion{H}{2} Regions}

\shorttitle{Masses, SFEs, and Dynamical Evolution}
\shortauthors{Pathak et al.}
\correspondingauthor{Debosmita Pathak}
% \email{pathak.89@buckeyemail.osu.edu}

\input{authors.tex}

% \suppressAffiliations

\begin{abstract}
We present measurements of the masses associated with $\sim18,000$ \ion{H}{2} regions across 19 nearby star-forming galaxies by combining data from JWST, HST, MUSE, ALMA, VLA, and MeerKAT from the multi-wavelength PHANGS survey. We report 10\,pc-scale measurements of the mass of young stars, ionized gas, \edit{and} older disk stars coincident with each \ion{H}{2} region, as well as the initial and current mass of molecular gas, atomic gas, and swept-up shell material\edit{, estimated from lower resolution data}. We find that the mass of older stars dominates over young stars at $\gtrsim10\rm\,pc$ scales, and ionized gas \edit{exceeds the stellar mass} in most optically bright \ion{H}{2} regions. Combining our mass measurements for a statistically large sample of \ion{H}{2} regions, we derive 10\,pc scale star-formation efficiencies $\approx6{-}17\%$ for individual \ion{H}{2} regions. Comparing each region's self-gravity with the ambient ISM pressure and total pressure from pre-supernova stellar feedback, we show that most optically bright \ion{H}{2} regions are over-pressured relative to their own self-gravity and the ambient ISM pressure, and that they are hence likely expanding into their surroundings. Larger \ion{H}{2} regions in galaxy centers approach dynamical equilibrium. The self-gravity of regions is expected to dominate over \edit{pre-supernova stellar feedback pressure at $\gtrsim130\rm\,pc$ and $60\rm\,pc$ scales} in galaxy disks and centers, respectively, \edit{but is always sub-dominant to the ambient ISM pressure on \ion{H}{2} region scales}. Our measurements have direct implications for the dynamical evolution of star-forming regions and the efficiency of stellar feedback in ionizing and clearing cold gas.
\end{abstract}


\section{Introduction}\label{sec:introduction}

\setcounter{footnote}{0} 
% --- WRITE INTRO ---
% What are HII regions and why are they interesting?
%Ionized gas or H$\alpha$ emission is one of the best tracers of recent ($\lesssim 10$ Myr old) star-formation in galaxies \citep[][]{2012KENNICUTT&EVANS}. Much of the H$\alpha$ emission in galaxies comes from $\sim10-100\rm\,pc$ scale regions of ionized gas, 
\ion{H}{2} regions trace the sites of recent massive star formation, created when young, massive OB stars ionize their surroundings. They have typical sizes of $\lesssim100\rm\,pc$, and can be easily identified by their strong nebular line (H$\alpha$) and IR dust emission \citep[e.g.,][]{2012KENNICUTT&EVANS}. 
% This emission offers powerful diagnostics of the physical state and metal content of the ionized gas as well as the nature of the stars powering the \ion{H}{2} region \citep[e.g.,][]{2019KEWLEY}. Moreover, 
Within \ion{H}{2} regions, young, massive stars exert multiple modes of feedback on the surrounding interstellar medium (ISM), including warm gas pressure, radiation pressure, stellar winds, and supernova feedback. As a result, \ion{H}{2} regions are key laboratories for studying star formation, stellar feedback, chemical enrichment, and dynamical evolution in the ISM \citep[e.g.,][]{2024SCHINNERER&LEROY}. 
%Understanding their structure, composition, and evolution is thus critical for both benchmarking theoretical models of \ion{H}{2} region evolution, and interpreting new extragalactic observations.

Recent high physical resolution ($\sim10{-}50\rm\,pc$) multi-wavelength observations have made it possible to build a comprehensive view of large samples of \ion{H}{2} regions beyond the Local Group \citep[e.g.,][]{2021BARNES, 2022SCHEUERMANN, 2022HANNON, 2022BARNES, 2024PEDRINI, 2025PATHAK} including $\sim 18,000$ regions with a detailed characterization of the stellar feedback properties \citep[][]{2025PATHAK}. These build on previous detailed studies of \ion{H}{2} regions in the Milky Way \citep[e.g.,][]{2020BARNES, 2021OLIVIER}, Magellanic Clouds \citep[e.g.,][]{2011LOPEZ, 2014LOPEZ}, and local galaxies \citep[$<5\rm\,Mpc$, e.g.,][]{2021LEVY, 2021MCLEOD, 2022DELLABRUNA, 2022COSENS}.

While feedback in \ion{H}{2} regions has been extensively studied, the masses of these star-forming regions have received comparatively less attention. A careful accounting of the mass of ionized, molecular, and atomic gas, as well as young and old stars is necessary to understand the impact of stellar feedback on the dynamical evolution of \ion{H}{2} regions and to estimate star formation efficiencies (SFE), i.e., the fraction of gas converted to stars. This Letter presents the first comprehensive inventory of the masses associated with a statistically large sample of \ion{H}{2} regions across nearby galaxies, with corresponding estimates of the SFE and dynamical state \citep[see e.g.,][for case studies with smaller samples]{1999CHURCHWELL&GOSS,2005RELANO,2022COSENS}.
%In parallel to feedback studies, a careful accounting of the local dynamical mass components associated with \ion{H}{2} regions is necessary to understand the actual impact of stellar feedback on region evolution and ISM dynamics.
% what has been comparatively less studied is the mass of ionized gas, mass of old stars coincident with regions
% for large samples of regions
% this has implications for the evolution of regions 
% highlight novelty for letter
%We estimate the young cluster mass, ionized gas mass, old (galactic disk) stellar mass, the initial molecular and atomic gas mass, and bounds on the dark matter and hot gas component associated with each region. 
% Our sample consists of $\sim$18,000 \ion{H}{2} regions covered by the multi-wavelength PHANGS (Physics at High Angular resolution in Nearby GalaxieS) surveys of nearby star-forming galaxies \citep[][]{2021LEROY,2022EMSELLEM, 2023LEE} at matched 10\,pc physical resolution (A. Barnes et al. submitted). 

For $\sim$18,000 \ion{H}{2} regions across 19 nearby galaxies, we present new 10\,pc-scale measurements of the mass of:
\begin{enumerate}[noitemsep]%[noitemsep, topsep=0pt]
    \item young stars (\S\ref{sec:estimating-MNew}),
    \item ionized gas (\S\ref{sec:estimating-MHII}),
    \item older stars from the galactic disk (\S\ref{sec:estimating-MOld}),
    \item initial molecular gas (\S\ref{sec:estimating-MMol}),
    \item additional disk components (initial atomic gas, and bounds on dark matter and hot X-ray emitting gas mass, \S\ref{sec:other-mass-comps}).
\end{enumerate}
We discuss the implications of our estimates for the efficiency of stellar feedback in ionizing and clearing local gas (\S\ref{sec:SFEs}), and the dynamical state of \ion{H}{2} regions (\S\ref{sec:region-dynamics}).
The analyzed \ion{H}{2} regions account for the bulk of ongoing star-formation in our galaxy sample \citep[e.g.,][]{2022BELFIORE,2022EMSELLEM}. Therefore, we expect their properties to be representative of \ion{H}{2} regions in normal star-forming galaxies in the $z\approx0$ universe.

\section{Estimating Mass Associated with \ion{H}{2} Regions} \label{sec:estimating-mass-comps}

\begin{figure*}[ht]
\centering
\includegraphics[width=1\textwidth]{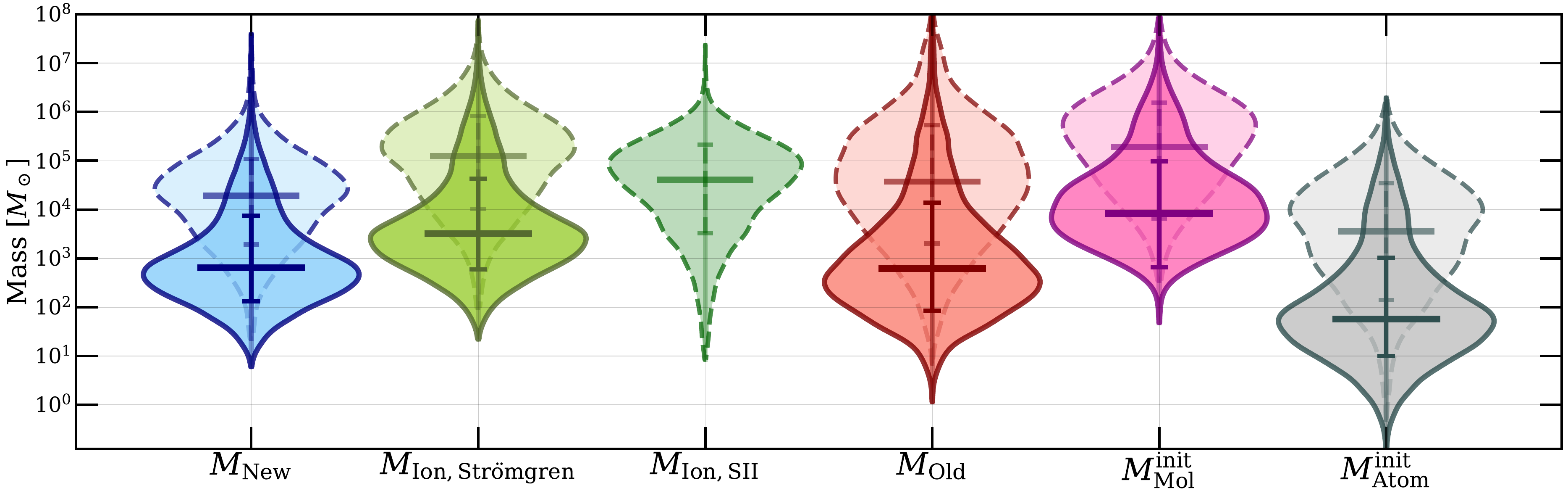}\\
\includegraphics[width=1\textwidth]{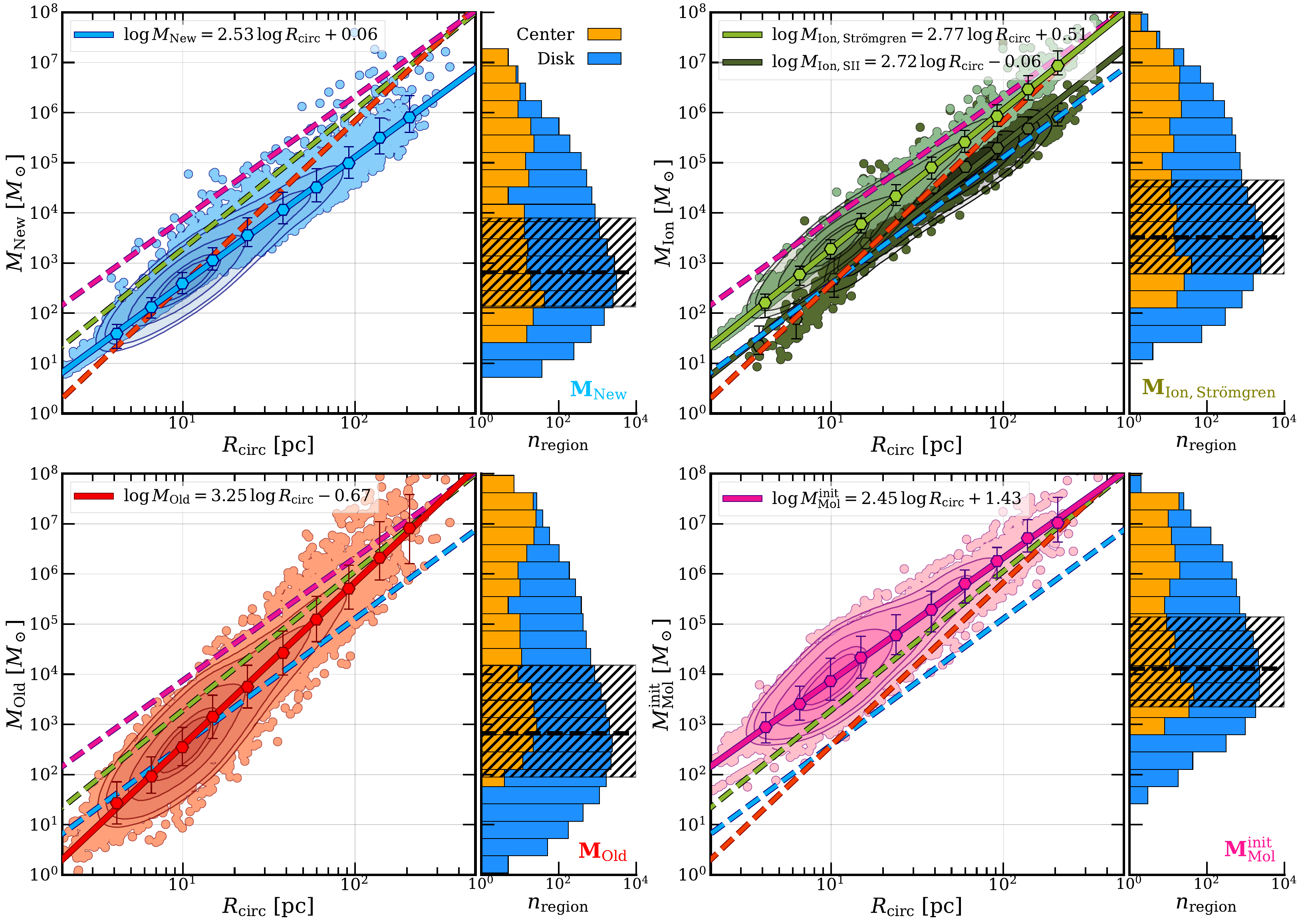}
\caption{The distribution of masses associated with our sample of $\sim18,000$ \ion{H}{2} regions (solid violins) and the subsample of $\sim 4,000$ regions where we can calculate [\ion{S}{2}] densities (dashed violins), including contribution from young clusters ($M_{\rm New}$, in blue, see \S \ref{sec:estimating-MNew}), ionized gas mass ($M_{\rm Ion}$; Strömgren in light green, [\ion{S}{2}]-emitting clump in dark green, see \S \ref{sec:estimating-MHII}), old stellar mass ($M_{\rm Old}$; red, see \S \ref{sec:estimating-MOld}), initial molecular gas mass ($M_{\rm Mol}^{\rm init}$ in pink, \S \ref{sec:estimating-MMol}), and initial atomic gas mass ($M_{\rm Atom}^{\rm init}$ in gray, \S\ref{sec:other-mass-comps}) as violin plots with the median and $16^{\rm th}{-}84^{\rm th}$ range included. Below we show how $M_{\rm New}$, $M_{\rm Ion}$, $M_{\rm Old}$, and $M_{\rm Mol}^{\rm init}$ scale with physical size ($R_{\rm circ}$). Contours show the $16^{\rm th}{/}25^{\rm th}{/}50^{\rm th}{/}75^{\rm th}{/}84^{\rm th}$ percentiles, and we include $\log-\log$ fits (bold lines, repeated as dashed lines in each panel) to the binned medians (hexagons), and the $16^{\rm th}{-}84^{\rm th}$ percentile scatter in each bin (error bars on hexagons). Attached to each scatter plot are stacked histograms of region masses for galaxy centers (orange) and disks (blue) \citep[following][]{2021QUEREJETA}; the sample median (dashed line), and $16^{\rm th}{-}84^{\rm th}$ percentile range (hatches). 
}
\label{fig:Mass-Terms-All4}
\end{figure*}

We estimate the masses associated with $\sim 18,000$ \ion{H}{2} regions across 19 nearby spiral galaxies \edit{that have been observed by} the PHANGS (Physics at High Angular resolution in Nearby GalaxieS) survey\edit{s} \citep[][]{2021LEROY,2022EMSELLEM, 2023LEE}, \edit{including coverage by H$\alpha$ imaging with} 10\,pc physical resolution \citep[A. Barnes et al. submitted][]{2025CHANDAR}.
% We estimate the masses associated with $\sim 18,000$ \ion{H}{2} regions across 19 nearby galaxies from the PHANGS-MUSE survey. %This includes the young cluster mass and mass in ionized gas using MUSE and HST narrowband H$\alpha$, the mass in old disk stars from JWST, and the molecular gas mass associated with each region using ALMA. 
The sample spans the star-forming main sequence, including 17 barred galaxies and two lower-mass spirals \citep[][]{2021LEROY, 2022EMSELLEM, 2023LEE, 2024WILLIAMS}. Our \ion{H}{2} regions are drawn from the PHANGS-MUSE nebular catalog \citep[][see also \citealt{2019KRECKEL,2022SANTORO}]{2023GROVES}, \edit{which includes} 17,615 Baldwin-Phillips-Terlevich (BPT)-selected \ion{H}{2} regions with joint MUSE, JWST, HST, and ALMA coverage \citep[same as][]{2025PATHAK}.

% For each \ion{H}{2} region we estimate the mass of the young cluster and ionized gas mass from MUSE and HST, mass of old stars from JWST, and bounds from ALMA on the initial molecular gas mass associated with each of the $\sim18,000$ \ion{H}{2} regions. The distributions for the full sample are summarized in Fig.~\ref{fig:Mass-Terms-All4}, \ref{fig:Mass-v-R-LHa}, and Table \ref{tab:summary-all}, estimated as follows.

% The region footprints in the MUSE catalogs are constructed by applying a modified version of \texttt{HIIPHOT} \citep[][]{2000THILKER} to the ``convolved and optimized'' H$\alpha$ emission line maps from \citet{2022EMSELLEM}. We then use the Baldwin-Phillips-Terlevich \citep[BPT;][]{1981BALDWIN} emission line diagnostic from the \citet{2023GROVES} catalog to select \ion{H}{2} regions and remove contaminants such as planetary nebulae, AGN, or supernova remnants. While we expect this selection to work relatively well, $\sim3\%$ of our BPT-selected \ion{H}{2} region sample overlap with supernova remnant candidates \citep[see, e.g.,][]{2024LI}.

\subsection{Current Masses}

\begin{figure*}[ht]
\centering
\includegraphics[width=1\textwidth]{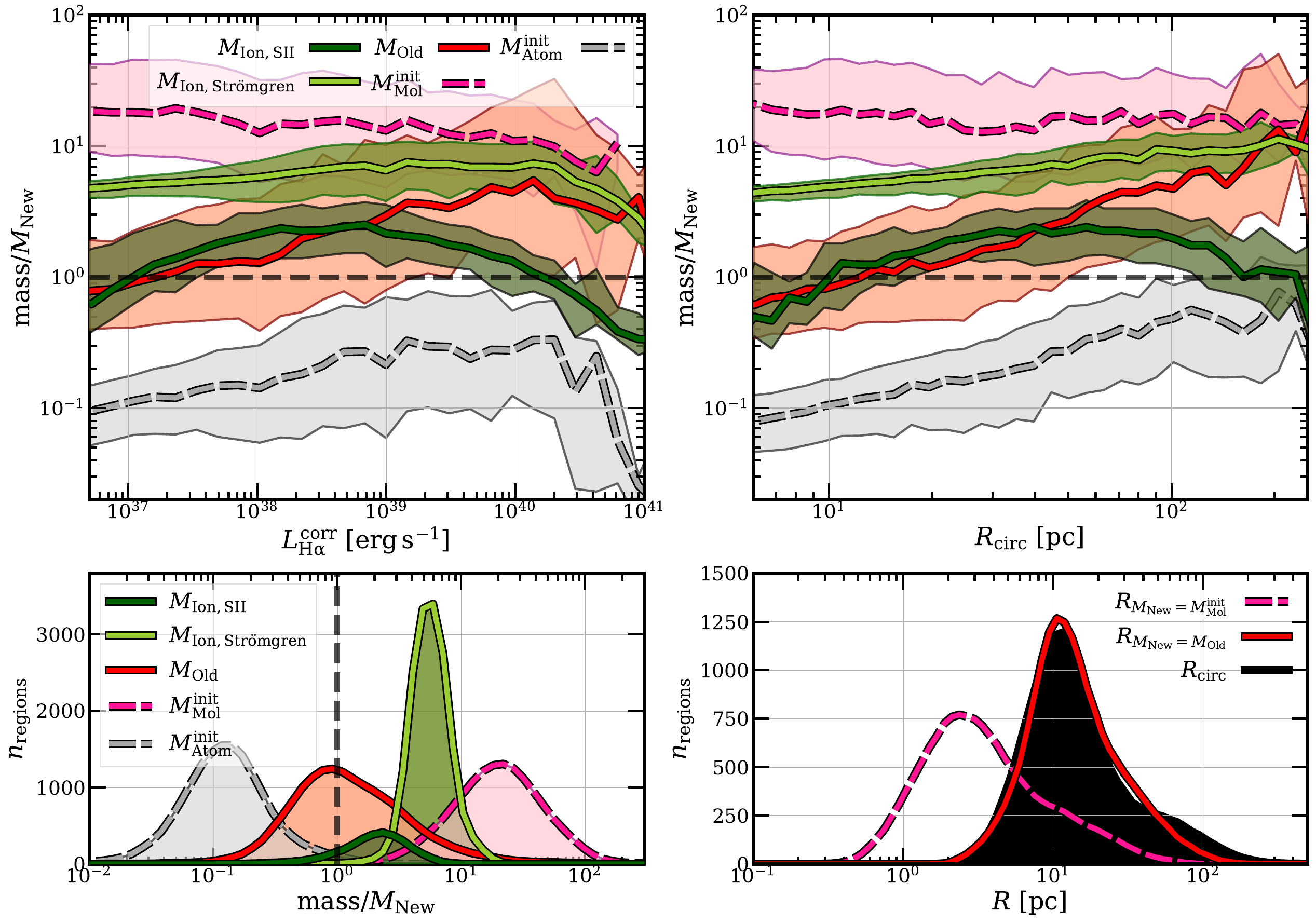}
\caption{Variation in the different mass components relative to the young cluster mass $M_{\rm New}$ for the full sample of \ion{H}{2} regions, binned in $L_{\rm H\alpha}^{\rm corr}$ (top left) and $R_{\rm circ}$ (top right), and summarized as histograms (bottom left). $M_{\rm Ion, SII}/M_{\rm New}$ (dark green) and $M_{\rm Ion, Str\ddot{o}mgren}/M_{\rm New}$ (light green), $M_{\rm Old}/M_{\rm New}$ (red), $M_{\rm Mol }^{\rm init}/M_{\rm New}$ (dashed pink), and $M_{\rm Atom}^{\rm init}/M_{\rm New}$ (dashed gray) are included for comparison. The dashed line indicates where a mass component is equal to $M_{\rm New}$. The bottom right panel includes the distribution of the measured $R_{\rm circ}$ (black), the estimated radius at which $M_{\rm New} = M_{\rm Mol}^{\rm init}$ (dashed pink), and radius where $M_{\rm New} = M_{\rm Old}$ (red).
}
\label{fig:Mass-v-R-LHa}
\end{figure*}

\subsubsection{Young Stars} \label{sec:estimating-MNew}

The mass of the young stars powering each region, $M_{\rm New}$, is estimated from its extinction-corrected H$\alpha$ luminosity, $L_{\rm H\alpha}^{\rm corr}$ \citep[][A. Barnes et al. submitted, for details see \citealt{2025PATHAK}]{2025CHANDAR}. As in \citet{2025PATHAK}, we use HST narrowband H$\alpha$ luminosities where available, and translate MUSE luminosities to be consistent with HST sizes otherwise \citep[see Appendix A in][for details]{2025PATHAK}. Extinctions are estimated from MUSE Balmer decrements \citep[][]{2022EMSELLEM, 2023GROVES, 2023BELFIORE}.

To translate the observed $L_{\rm H\alpha}^{\rm corr}$ to an associated young stellar mass, we calculate the bolometric luminosity-weighted average H-ionizing luminosity over the first 4 Myr \citep[before the ionizing photon flux drops and supernovae go off, see][]{2021BARNES} of simple stellar population (SSP) evolution with the default \textsc{Starburst99} \citep{1999LEITHERER,2014LEITHERER} parameters\footnote{Run for a $10^6 M_\odot$ cluster with fully populated \citet{2001KROUPA} initial mass function (IMF) with a maximum stellar mass of $100 \, M_\odot$, at solar metallicity, and for non-rotating stars.} 
%This yields $L_{\rm bol} \approx 88 \times L_{\rm H \alpha}^{\rm corr}$, consistent with \citet{2021BARNES}. 
which we then compare to the ionizing luminosity of each region \citep[consistent with][]{2021BARNES, 2025PATHAK}. This yields $M_{\rm New}/L_{\rm H\alpha}^{\rm SB99} = 0.112 M_\odot / L_\odot$.\footnote{For a fully sampled IMF, the conversion from $L_{\rm H\alpha}$ to $M_{\rm New}$ depends on the age of the stellar population, with commonly used values of $M_{\rm New}/L_{\rm H\alpha}^{\rm SB99}$ ranging from $\approx 0.08{-}0.18 \, M_\odot / L_\odot$ \citep[see discussion in][]{2025PATHAK}.}

We correct the observed $L_{\rm H\alpha}^{\rm corr}$ to take into account the fraction of ionizing photons escaping the region, $f_{\rm esc}$, as, $L_{\rm H\alpha}^{\rm corr}/(1-f_{\rm esc})$.
% Since some fraction $f_{\rm esc}$ of the ionizing photons escape the region and contribute to the diffuse ionized gas (DIG), we apply a correction factor to the estimate of the total H$\alpha$ luminosity associated with the region, $L_{\rm H\alpha}^{\rm corr}/(1-f_{\rm esc})$. 
\edit{Theory predicts large variations in $f_{\rm esc}$ due to environment and geometry \citep[e.g.,][]{2018HOWARD, 2019JKIM, 2020HE, 2022KIMM, 2025MENON}.}
While mean $f_{\rm esc}\sim30\pm 10\%$ for \ion{H}{2} regions in normal star-forming galaxies, $f_{\rm esc}$ for any individual region is highly uncertain \citep[e.g.,][A. Barnes et al. submitted]{2022BELFIORE, 2022SCHEUERMANN}. 
We thus assume a fixed $f_{\rm esc} = 0.3$ and estimate $M_{\rm New}$ as,
\begin{equation}
\begin{split}
    M_{\rm New} &= \dfrac{0.112 M_\odot}{L_\odot} \times \dfrac{L_{\rm H\alpha}^{\rm corr}}{(1-f_{\rm esc})}\\
    &= 160 \biggl ( \dfrac{L_{\rm H\alpha}^{\rm corr}}{10^3 L_\odot} \biggr ) \biggl ( \dfrac{1-f_{\rm esc}}{0.7} \biggr )^{-1}  M_\odot .
\end{split}\label{eq:MNew-scaled}
\end{equation}
%Fig.~\ref{fig:Mass-Terms-All4} summarizes the variation in $M_{\rm New}$ with $R_{\rm circ}$ and by local environment for our sample of $\sim18,000$ \ion{H}{2} regions.  

Note that Eq.~\eqref{eq:MNew-scaled} does not account for the $E>24.5 \, \rm eV$ photons absorbed by helium ($\sim5-10\%$ level effect) \edit{or} ionizing photons that are absorbed by dust before H-ionization, \edit{which lead to a likely under-estimate of the total mass of young stars.} \edit{Absorption by dust can be significant}, especially in the massive and dusty \ion{H}{2} regions in galaxy centers \edit{\citep[$\sim30\%$, see][or $\sim 10-50\%$, see \citealt{2019JKIM, 2025MENON}]{2018BINDER}}. However, most regions in our sample show relatively low $A_{\rm V}\approx0.6{-}1$~mag \citep[see][]{2025PATHAK}, where this fraction is expected to be much lower, but remains unconstrained.
%, the details of which will be studied in the future with higher resolution Paschen-$\alpha$ extinctions. 

\edit{We likely underestimate $M_{\rm New}$ for most clusters at low-$L_{\rm H\alpha}^{\rm corr}$, since our conversion is based on SSP models that are non-stochastic \citep[see e.g.,][for stochastic IMF sampling]{2012DASILVA} and do not include binaries \citep[e.g.,][]{2020GOTBERG}.} Binaries may cause additional ionization at ages $\gtrsim 10$~Myr. It is uncertain to what degree these populations affect \ion{H}{2} regions, but both topics are expected to be the subject of follow-up work in PHANGS.

% Note that our conversion is based on SSP models that are not non-stochastic and do not include binaries. Because we do not account for stochastic sampling of the IMF, we likely underestimate $M_{\rm New}$ for most clusters at low-$L_{\rm H\alpha}^{\rm corr}$. Binaries lead to additional ionizing photon production, and can dominate at intermediate ages $\gtrsim 10$~Myr. The degree to which such populations contribute to \ion{H}{2} regions remains uncertain. Both topics are expected to be the subject of follow-up work in PHANGS.

%ollow-up work will quantify the variation due to stochasticity and binarity, which becomes relevant for low-mass clusters.
% \textcolor{red}{Add caveat: stochasticity and the likely frequent resulting underestimate of $M_{\rm new}$ from H$\alpha$ for low end stellar masses (from Aida). It is also worth mentioning the assumption of solar metallicity and of non-rotating single stars in the calculation of the spectrum of the star cluster. Can you speculate how changing these would influence your conclusions? (Ralf)}

\subsubsection{Ionized Gas} \label{sec:estimating-MHII}

The ionized gas mass can be determined from the H$\alpha$ luminosity assuming Case B recombination with a given volume density. We estimate the ionized gas density in two ways: (1) using the Strömgren sphere approximation and (2) \edit{from} the [\ion{S}{2}] doublet ratio. These yield upper and lower bounds on the total ionized gas mass, respectively.

\edit{For a} smooth spherical gas distribution \edit{(clumping factor $f_{\rm c} = 1$)} with a fixed density, \edit{the Strömgren calculation yields,} 
\begin{equation}
    n_{\rm H, Str\ddot{o}mgren} = \sqrt{\dfrac{3Q_0 }{4 \pi \alpha_{\rm B} R_{\rm circ}^3 \edit{f_{\rm e} f_{\rm c}}} }, \label{eq:Stromgren-nH}
\end{equation}
where $1.367 Q_0 / \rm s^{-1} = 10^{12} L_{\rm H\alpha}^{\rm corr} / (\rm erg \, s^{-1})$ \citep[][]{2006OSTERBROCK-book}, the factor of \edit{$f_{\rm e} = 1.08$} accounts for electrons from singly ionized He in addition to H at \edit{an abundance representative of the ISM of external galaxies \citep[as in][]{2013BOLATTO}, assuming $M_{\rm He}+M_{\rm H} = 1.36 M_{\rm H}$ or $n_{\rm He}/n_{\rm H} = 0.08$}. $\alpha_{\rm B} = 3.1\times 10^{-13} ~{\rm cm^3 \, s^{-1}}$ is the Case B H-recombination coefficient at electron temperature $T_{\rm e}=8,000 \rm \, K$ \citep[][]{1995STOREY&HUMMER} representative of the PHANGS-MUSE sample \citep[][]{2023EGOROV}. $R_{\rm circ}$ is the circularized radius corresponding to an isophotal area measured from 10\,pc-resolution HST narrowband H$\alpha$ imaging (see A. Barnes et al. submitted). The corresponding \ion{H}{2} and ionized gas mass of an \ion{H}{2} region is then 
\begin{equation}
\begin{split}
M_{\rm HII, Str\ddot{o}mgren} &= m_{\rm H} \, n_{\rm H, Str\ddot{o}mgren} \dfrac{4}{3}\pi R_{\rm circ}^3 , \\ 
M_{\rm Ion, Str\ddot{o}mgren} &= 1.36 \, M_{\rm HII, Str\ddot{o}mgren} \\ &= 1.2\times10^3 \biggl ( \dfrac{L_{\rm H\alpha}^{\rm corr}}{10^3 L_\odot} \biggr )^{1/2} \biggl ( \dfrac{R_{\rm circ}}{10 \rm \, pc} \biggr )^{3/2} M_\odot.\label{eq:estimating-MHII-Stromgren}
\end{split}
\end{equation}
\edit{Two caveats are worth bearing in mind. First, there is some ambiguity regarding the appropriate size. We use $R_{\rm circ}$, based on the area above a fixed H$\alpha$ intensity threshold in the HST maps, which matches the region used to derive cluster and gas properties. However, this threshold captures varying fractions of the total H$\alpha$ flux—e.g., going down in sensitivity to $\sim$10\% of the peak value in bright regions (and therefore a large fraction of the total flux) as compared to only down to $\sim$50\% in faint \ion{H}{2} regions (and therefore a smaller fraction of the total flux) 
% \edit{it may represent $\sim 10\%$ of the peak in bright regions (and therefore a large fraction of the total flux) and $\sim 50\%$ in faint ones (and therefore a smaller fraction of the total flux). 
Second, Eq.~\eqref{eq:Stromgren-nH} assumes a uniform volume density ($f_{\rm c} = 1$) while the actual ionized gas distribution is generally more clumpy \citep[$f_{\rm c}>1$; see e.g.,][]{2025LANCASTER}.  As a result the Strömgren estimate represents a \textit{lower} bound on the actual physical density $n_{\rm H}$ of the clumpy H$\alpha$-emitting gas, but an \textit{upper} bound on the volume-average ionized gas density $\langle n_{\rm H} \rangle$, and hence an upper bound on the total ionized gas mass.}

%As shown in Fig.~\ref{fig:Mass-v-R-LHa} and Fig.~\ref{fig:Mass-Terms-All4}, $M_{\rm Ion, Str\ddot{o}mgren}>M_{\rm New}$ at all $R_{\rm circ}$.

We also use the [\ion{S}{2}] line ratio to estimate the electron number density $n_{\rm e, SII}$ using \textsc{pyneb} \citep[][]{2015LURIDIANA} for a subset of $3,221$ regions where the [\ion{S}{2}] ratio differs from the low density limit at high confidence \citep[see also][and A. Barnes et al. submitted]{2021BARNES}.
Assuming that the H$\alpha$ emission comes from the same volume probed by the [\ion{S}{2}] doublet ($V_{\rm SII}$), we balance the observed H$\alpha$ luminosity against the recombination rate \citep[][]{2006OSTERBROCK-book} suggested by $n_{\rm e, SII}$: 
%and is not necessarily representative of the volume-average density, we use the observed luminosity of each region to constrain the \ion{S}{2}-emitting volume $V_{\rm SII}$ within each \ion{H}{2} region as
\begin{equation}
\begin{split}
L_{\rm H\alpha}^{\rm corr} &= 0.45 \langle h \nu\rangle_{\rm H\alpha} n_{\rm e}  n_{\rm H} \alpha_B V_{\rm SII} \\&= 0.45 \langle h \nu\rangle_{\rm H\alpha}  \dfrac{n_{\rm e,SII}^2}{f_{\rm e}} \alpha_B V_{\rm SII} .
\end{split}
\end{equation}
Then, 
%yielding a corresponding ionized gas mass of
% can do eqnarray
\begin{equation}
\begin{split}
M_{\rm HII, SII} &= m_{\rm H} \, \dfrac{n_{\rm e, SII}}{f_{\rm e}} V_{\rm SII} , \\ 
M_{\rm Ion, SII} &= 1.36 M_{\rm HII, SII} \\ 
&= 2.1\times10^2 \biggl ( \dfrac{L_{\rm H\alpha}^{\rm corr}}{10^3 L_\odot} \biggr ) \biggl ( \dfrac{n_{\rm e, SII}}{50 \rm \, cm^{-3}} \biggr )^{-1} M_\odot. \label{eq:estimating-MHII-SII}
\end{split}
\end{equation}
%$M_{\rm HII, SII}$ represents a lower bound on the total ionized gas mass since the \ion{S}{2} density diagnostic is only sensitive to $\sim50{-}10^4 \rm\,cm^{-2}$ ionized gas, and is not representative of lower-density volume-filling ionized gas. 
%It is important to bear in mind that the $\sim3,000$ regions where we can estimate $n_{\rm e}$ from the [\ion{S}{2}] doublet are biased towards the densest regions, which also tend to be the most luminous, since the [\ion{S}{2}] density diagnostic is only sensitive to $\approx50{-}10^4 \rm\,cm^{-3}$ ionized gas, and is not representative of lower-density and more volume-filling ionized gas.

%As summarized in Fig.~\ref{fig:Mass-Terms-All4} and \ref{fig:Mass-v-R-LHa}, $M_{\rm HII, SII}\approx 2\times M_{\rm New}$ for most regions with $L_{\rm H\alpha}^{\rm corr} \approx 10^{38}{-}10^{40}\rm\,erg\,s^{-1}$ or $R_{\rm circ}\approx10{-100}\rm\,pc$. 
\edit{Since [\ion{S}{2}] emission increases with density, $n_{\rm e, SII}$ traces the densest clumps of ionized gas.
% Since $n_{\rm e, SII}$ is sensitive to the clumps with the highest [\ion{S}{2}] luminosities, 
We expect this to represent an upper limit to the physical density and a lower limit to the total mass \citep[see][for details of using MUSE density diagnostics]{2023EGOROV}. }
% $R_{\rm circ}$ expresses the area of the region above a fixed intensity threshold in the HST H$\alpha$ map. These have the advantage of capturing exactly the area of the region used to calculate the young cluster and ionized gas properties from HST images. However, for regions of different brightness, a fixed isophote captures different fractions of the total H$\alpha$ flux associated with the region. That is, this threshold may represent 10\% of the peak in a brighter region and 50\% of the peak in a fainter region. 
\edit{Finally,} the [\ion{S}{2}] doublet enters the low density limit for many of our regions \citep[see][]{2021BARNES}, biasing \edit{the sample where} $n_{\rm e, SII}$ estimates \edit{are possible} towards the densest regions.
%so that the $\sim3,000$ regions where we can estimate $n_{\rm e}$ from the [\ion{S}{2}] doublet are biased towards the densest regions.

\subsubsection{Older Stars in the Galactic Disk} \label{sec:estimating-MOld}

The \ion{H}{2} regions in nearby disk galaxies evolve within a massive, extended galactic disk of older ($\gg 10 \rm \, Myr$) stars that contribute to the mass enclosed within the \ion{H}{2} region. We estimate this contribution using inclination-corrected JWST/NIRCam F300M ($3\,\mu$m) intensities \citep[][for details on data processing, see \citealt{2024WILLIAMS}]{2023LEE}. 

We calculate the F300M surface brightness profile of the galaxies, and evaluate its value $I_{3 \rm \mu m}$ at the galactic radius of each region 
% To do this, we calculate radial profiles of F300M surface brightness, $I_{3 \rm \mu m}$. We then evaluate the azimuthally averaged $3\mu$m surface brightness at the galactocentric radius of each region 
to estimate the stellar mass surface density of disk stars, $\Sigma_{\rm Old}$, via
\begin{equation}
\label{eq:oldstars}
    \dfrac{\Sigma_{\rm Old}}{\rm M_\odot \, pc^{-2}} = 2.6 \times 10^2 \biggl ( \dfrac{\Upsilon_*}{0.5} \biggr ) \biggl ( \dfrac{I_{3 \rm \mu m}}{\rm 1 \, MJy \, sr^{-1}} \biggr ),
\end{equation}
%\times \cos{i}
where $\Upsilon_*$ is the mass-to-light ratio ($M/L$) in solar units. We adopt the specific star formation rate-dependent near-IR $\Upsilon_*$ from \citet{2019LEROY}, implemented in \citet{2022SUN} (see those papers for more details), \edit{which shows good consistency with MUSE stellar mass surface densities \citep[][]{2022EMSELLEM}.} Compared to \citet{2019LEROY}, which utilized $3.4\mu$m emission, we adjust the pre-factor in Eq.~\ref{eq:oldstars} assuming that the Sun's spectrum follows a Rayleigh Jeans distribution $L_\nu \propto \nu^2$ near $3\mu$m. % what I have looks consistent with Willott+2018
%As in \citet{2019LEROY}, we first estimate the solar luminosity at $3\,\mu$m,
%\begin{equation}
%   \nu L_{\rm \nu , \odot}^{\rm 3 \mu m} \approx 2.2631 \times 10^{32} {\rm \, erg \, s^{-1}} \approx 0.059 L_\odot,
%\end{equation}
%which 
Here we use the radial profile rather than evaluating $I_{\rm 3\mu m}$ at the location of the region to avoid contamination of this estimate by the young stellar population. \citet{2021MEIDT,2024QUEREJETA} show that azimuthal variations in stellar structure are relatively modest ($\lesssim 20{-}30\%$) in these targets.

%the azimuthal smoothness of the total stellar mass profile of each galaxy to estimate $\Sigma_{\rm Old}$ of the background stars. We choose this approach as opposed to $3\,\mu$m aperture photometry targeting \ion{H}{2} regions because the latter will include significant emission from the young stellar population, which will also have a low and uncertain mass-to-light ratio.
%, and difficulty in constraining the mass-to-light ratio within nebular regions. 

Assuming a vertical scale-height of the stellar disk $h^{\rm exp}_{z} = L_{\rm *}/7.3$ \citep[for an exponential vertical density profile, as in][]{2002KREGEL}\footnote{Since there is no consensus on the functional form of the vertical stellar distribution in galaxies, we adopt the exponential form which is in slightly better agreement with observations that often show profiles more centrally concentrated than ${\rm sech}^2$ \citep[e.g.,][]{2020DOBBIE,2025JOG}. Adopting a ${\rm sech}^2$ profile instead \citep[as in][]{2015SALO} would predict 0.5 times the $\rho_{\rm Old}$ at the mid-plane, although both profiles agree at large $z$ \citep[see][]{2020SUN}.}
%the measurements of \citet{2002KREGEL} and \citet{2015SALO} appear to substantially agree on the shape of the disk at large $z$. 
where $L_{\rm *}$ is the exponential scale-length of the disk, we derive the average background stellar mass density,
\begin{equation}
\begin{split}
\rho_{\rm Old} &= \dfrac{\Sigma_{\rm Old}}{2h^{\rm exp}_{z}} , \\ 
\dfrac{\rho_{\rm Old}}{\rm M_\odot \, pc^{-3}} &= 0.52 \biggl ( \dfrac{\Upsilon_*}{0.5} \biggr ) \biggl ( \dfrac{I_{3 \rm \mu m}}{\rm 1 \, MJy \, sr^{-1}} \biggr ) \biggl ( \dfrac{2 h^{\rm exp}_{z}}{500 \, \rm pc} \biggr )^{-1}, \label{eq:estimating-MOld}
\end{split}
\end{equation}
%\times \cos{i}
at the galactocentric radius of each \ion{H}{2} region. We record $\rho_{\rm Old}$ in Table \ref{tab:summary-all}. Given this density and a volume, with the assumption that \ion{H}{2} regions lie close to the mid-plane (within $2 h^{\rm exp}_{z}$), we can estimate the mass ($M_{\rm Old}$) of disk stars. For our adopted footprint size, $R_{\rm circ}$, we calculate $M_{\rm Old}$ within each region as,
\begin{equation}
\begin{split} \label{eq:estimating-MOld-final}
M_{\rm Old} &= \rho_{\rm Old} \times \dfrac{4}{3} \pi R_{\rm circ}^3. %\\ 
%\dfrac{M_{\rm Old}}{M_\odot} &= 737 \biggl ( \dfrac{\Upsilon_*^{3.4 \rm \mu m}}{0.5} \biggr ) \biggl ( \dfrac{I_{3 \rm \mu m}}{\rm 1 \, MJy \, sr^{-1}} \biggr ) \\
%&\quad\quad\quad\quad\quad \times \biggl ( \dfrac{h_{\rm disk}}{1.5 \, \rm kpc} \biggr )^{-1} \biggl ( \dfrac{R_{\rm circ}}{10 \,\rm pc} \biggr )^3.
\end{split}
\end{equation}
%\times \cos{i}
%
%From Fig.~\ref{fig:Mass-Terms-All4} and \ref{fig:Mass-v-R-LHa}, $M_{\rm Old}>M_{\rm New}$ at $R_{\rm circ}\gtrsim40\rm\,pc$ scales on average, further discussed in \S\ref{sec:what-scales-do-mass-comps-dominate}.

% \textcolor{red}{Atomic gas - similar to $\rho_*$ you are fundamentally estimating density + a size scale. But $\rho_{\rm HI}$ is less than $\rho_*$ - you might cast these three densities together as large scale disk densities.}

\subsection{Initial Gas Masses}

\subsubsection{Molecular Gas} \label{sec:estimating-MMol}

% \textcolor{red}{You do TWO estimates right? Be clear with a couple sentences here about the issue (feedback changes initial distribution, plus resolution iffy. Then say we do two things: measure AT the site of the current region to get CURRENT and take a typical regional mass-weighted Sigma\_mol and scale by the size of the HII region to get INITIAL.} ---> DP: we talked about this, I'll add current MMol to the SFE section, but only focus on the initial MMol here

We also estimate the initial molecular gas mass associated with each \ion{H}{2} region before star-formation. We use the mass-weighted average molecular gas surface density \citep[using $\alpha_{\rm CO}$ prescription from][]{2024SCHINNERER&LEROY} at $150$\,pc resolution, $\left< \Sigma_{\rm Mol} \right>_{150}$, within the 1.5\,kpc diameter hexagonal region containing that region calculated by \citet{2022SUN}. This provides an estimate of the typical surface density of molecular clouds in the same part of the galaxy as the target region, and we expect that it represents a reasonable estimate of the initial conditions before feedback.
%within the local 1.5 kpc from the mega tables compiled by \citet{2020SUN}. 
% Since the local average $\Sigma_{\rm Mol}$ may still be an underestimate of the initial molecular gas density, we also use the mass-weighted local average density $\Sigma_{\rm \langle Mol \rangle}$ to provide an upper bound. 
% Since the molecular gas scale height is not precisely known and is likely to be of the same order as the \ion{H}{2} region sizes, we directly scale the local molecular gas surface density by the area of each \ion{H}{2} region footprint to estimate the initial molecular gas mass as,
We assume the molecular gas scale height to be of the same order as the \ion{H}{2} region sizes. Therefore, we estimate the initial molecular gas mass from the area of each \ion{H}{2} region footprint,
\begin{equation}
\begin{split}
M_{\rm Mol}^{\rm init} &= 1.36 \, \left< \Sigma_{\rm H_2} \right>_{150} \, \pi R_{\rm circ}^2 =  \left< \Sigma_{\rm Mol} \right>_{150} \, \pi R_{\rm circ}^2, \\ 
\dfrac{M_{\rm Mol}^{\rm init}}{M_\odot} &= 3.1\times 10^3 \biggl ( \dfrac{\left< \Sigma_{\rm Mol} \right>_{150}}{10 \rm \, M_\odot pc^{-2}} \biggr ) \biggl ( \dfrac{R_{\rm circ}}{10 \,\rm pc} \biggr )^2.
\end{split} \label{eq:estimating-MMol}
\end{equation}
%$M_{\rm Mol}^{\rm init}$ is consistently higher than all the other major mass terms. 
% Note that $M_{\rm H_{2}} = M_{\rm Mol}/1.36$. 

Note that the 150~pc resolution used here is larger than the the size of the \ion{H}{2} regions. $\left< \Sigma_{\rm Mol} \right>_{\rm 150}$ \edit{is predictive of the surface densities at higher resolution} ($120$, $90$, and $60$~pc) well, but due to clumping, the higher resolutions show higher $\left< \Sigma_{\rm Mol} \right>$ \citep{2018SUN,2022SUN,2025LEROY}; for example, $\left< \Sigma_{\rm Mol} \right>_{\rm 60}$ is $\approx 0.13$~dex or $1.35 \times$ higher than $\left< \Sigma_{\rm Mol} \right>_{\rm 150}$ in \citet{2025LEROY}. Based on this, we \edit{might expect our $M_{\rm Mol}$ to be underestimated by $\sim 25{-}50\%$.} %Overall, caution that predicting $10$~pc from $150$~pc surface densities represents a significant extrapolation so that $M_{\rm mol}^{\rm init}$ is correspondingly uncertain}.

Our $M_{\rm Mol}^{\rm init}$ thus represents a statistical estimate of the original mass. We also measure the \textit{current} $\Sigma_{\rm Mol}$ at the location of each region. Although we do not expect that there is significant molecular gas within the \ion{H}{2} regions themselves, this measurement captures any molecular gas near the region. Contrasting the current 150\,pc associated and estimated initial $\left< \Sigma_{\rm Mol} \right>_{\rm 150}$ also provides a statistical constraint on how much local gas is ionized and cleared by feedback (see \S\ref{sec:SFEs}).

\subsubsection{Additional Disk Components} \label{sec:other-mass-comps}

In addition to stars, galaxy disks contain atomic gas and dark matter. Although we expect \ion{H}{1} within the \ion{H}{2} region to be dispersed or ionized we estimate the likely initial atomic gas mass of each region based on lower resolution 21-cm \ion{H}{1} mapping from MeerKAT \citep[IC~5332, NGC~1300, NGC~1512, NGC~1566, NGC~1672, NGC~4535, NGC~5068, and NGC~7496][and D. J. Pisano et al. in preparation]{2024DEBLOK,2024EIBENSTEINER} and the VLA \citep[NGC~0628, NGC~1087, NGC~1385, NGC~3351, NGC~2835, NGC~3627, NGC~4254, NGC~4303, NGC~4321][]{2008WALTER,2009CHUNG,2024CHIANG}. No \ion{H}{1} data were available for NGC~1365 and NGC~1433. 
%This accounts for a subset of 16,549 \ion{H}{2} regions from 17 galaxies where 
We estimate the \ion{H}{1} column density towards each region at 2\,kpc resolution from the inclination-corrected integrated intensity,
\begin{equation}
    \dfrac{N(\rm HI)}{\rm cm^{-2}} = 1.823\times10^{18} \dfrac{I_{\rm HI}}{\rm K \, km\, s^{-1}},
\end{equation}
and the corresponding midplane initial atomic gas density assuming the \ion{H}{1} is optically thin, $\rho_{\rm Atom} = 1.36 \, m_{\rm HI}N(\rm HI)/h_{\rm HI}^{\rm FWHM}$, where $h_{\rm HI}^{\rm FWHM}$ is the vertical FWHM of the \ion{H}{1} gas disk (generally $h_{\rm HI}^{\rm FWHM} > 2h_{z}^{\rm exp}$). We set $h_{\rm HI}^{\rm FWHM} = 1\rm\,kpc$, typical for the optical disk of star-forming massive spiral galaxies \citep[see e.g.,][]{2021RANDRIAMAMPANDRY, 2022ZHENG}.
The initial atomic gas mass $M_{\rm Atom}^{\rm init}$ is then (similar to Eq.~\eqref{eq:estimating-MOld}), 
\begin{equation}
\begin{split} % add eq for I21cm to SigmaAtom
M_{\rm Atom}^{\rm init} &= \rho_{\rm Atom} \times \dfrac{4}{3} \pi R_{\rm circ}^3, \\ 
\dfrac{M_{\rm Atom}^{\rm init}}{M_\odot} &= 41.6 \biggl ( \dfrac{I_{\rm HI} }{500 \rm \, K \, km \, s^{-1}} \biggr ) \biggl ( \dfrac{R_{\rm circ}}{10 \,\rm pc} \biggr )^3.
\end{split} \label{eq:estimating-MAtom}
\end{equation}
$M_{\rm Atom}^{\rm init}$ is roughly two orders of magnitude lower than $M_{\rm Mol}^{\rm init}$ (Fig.~\ref{fig:Mass-v-R-LHa} and Table \ref{tab:summary-short}). 

To estimate the dark matter contribution to the dynamical mass of \ion{H}{2} regions, we compare the typical dark matter density $\rho_{\rm DM}$ between $1{-}10\rm\,kpc$ with the baryonic components estimated so far. For a subset of our targets, recent observational constraints on their dark matter density profiles are available from \citet{2025VIJAYAKUMAR}. This work suggests $\rho_{\rm DM} \approx 10^{-1} {-} 10^{-2} \;M_\odot\rm\,pc^{-3}$ at $R_{\rm gal}=1$~kpc and $\approx 10^{-2} {-} 10^{-3}\;M_\odot\rm\,pc^{-3}$ at $R_{\rm gal}=10$~kpc, within which 95\% of our \ion{H}{2} regions lie. This is consistent with simulation predictions for massive star forming halos
%where the typical $\rho_{\rm DM}$ ranges from roughly $10^{-2} M_\odot\rm\,pc^{-3}$ in galaxy centers to $10^{-3} M_\odot\rm\,pc^{-3}$ at $R_{\rm gal} = 10\rm\, kpc$ 
\citep[see e.g.,][]{2014DICINTIO, 2023JIANG}. Given that these numbers are roughly one order of magnitude lower than $\rho_{\rm Atom}$ (see Table \ref{tab:summary-all}), we do not expect dark matter to contribute appreciably to the total mass of \ion{H}{2} regions.  

% hot Xray gas from Laura
Finally, the average density of X-ray emitting hot gas in galaxy disks is $n_{\rm e, X} \approx 10^{-3} {-} 10^{-2} \, \rm cm^{-3}$ \citep[see e.g.,][]{2012MINEO}, which translates to $\rho_{\rm X} \approx 3\times10^{-5} {-} 3\times 10^{-4} \rm \, M_{\odot} \, pc^{-3}$, and therefore negligible.
% which is roughly $2{-}3$ orders of magnitude lower than even $\rho_{\rm Atom}^{\rm init}$, and can be ignored for mass estimates.

\begin{figure*}[t]
\centering
\includegraphics[width=1\textwidth]{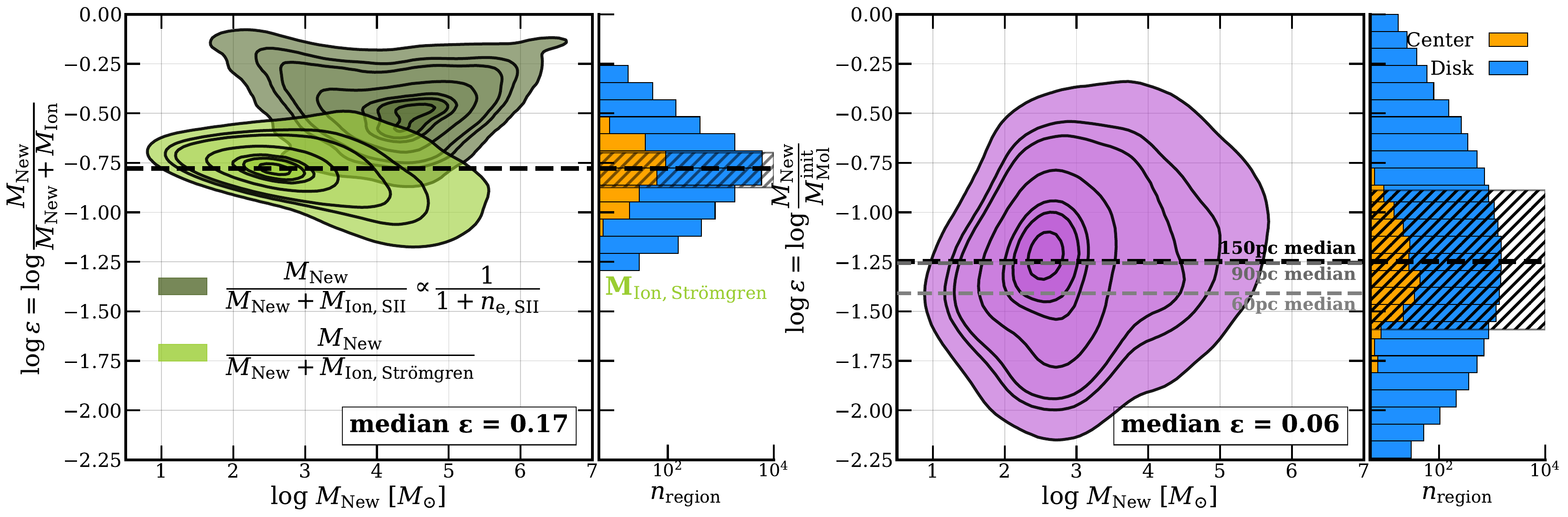} \\
\includegraphics[width=1\textwidth]{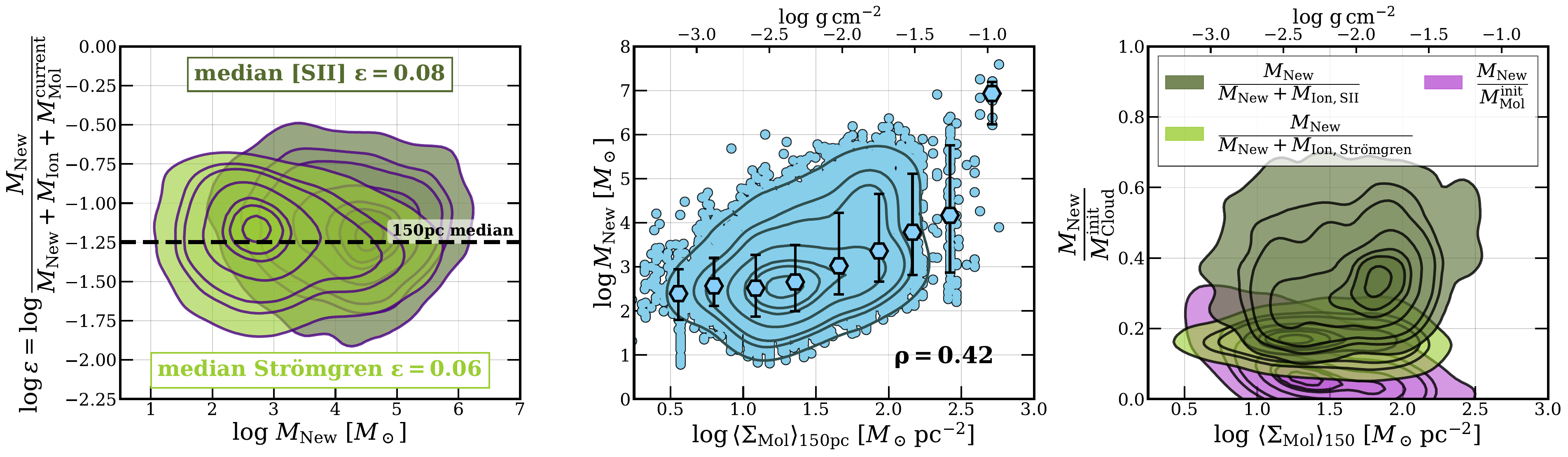} \\
\includegraphics[width=1\textwidth]{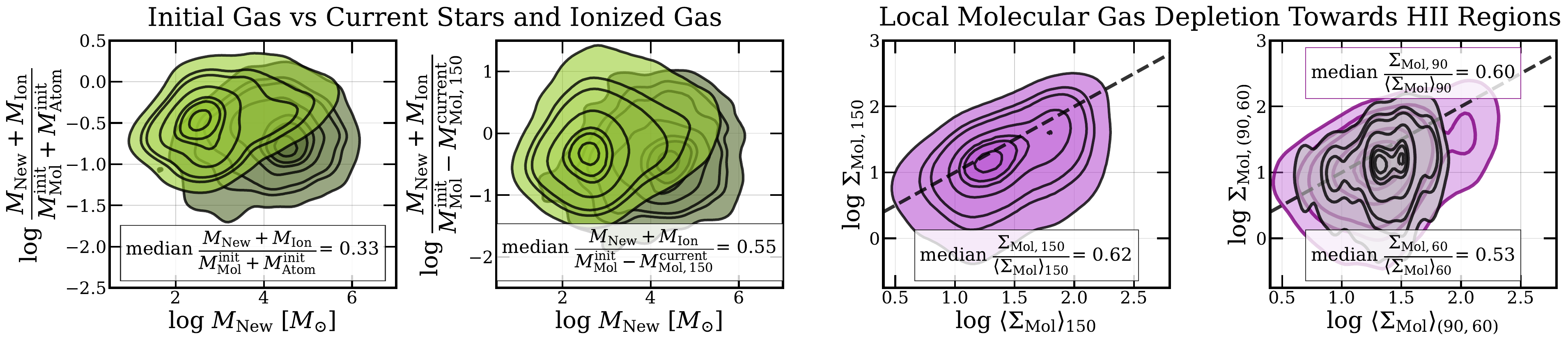} 
\caption{\textbf{Top:} Summarizing \ion{H}{2} region-scale SFEs. Data density contours for the $5^{\rm th}{/}16^{\rm th}{/}25^{\rm th}{/}50^{\rm th}{/}75^{\rm th}{/}84^{\rm th}{/}95^{\rm th}$ percentile SFE using $M_{\rm Ion, SII}$ (dark green), $M_{\rm Ion, Str\ddot{o}mgren}$ (light green), and $M_{\rm Mol}^{\rm init}$ (purple). The full histogram and medians as before (dashed line, printed) for $M_{\rm Ion, Str\ddot{o}mgren}$ and $M_{\rm Mol}^{\rm init}$ are included. We also include the medians using $M_{\rm Mol}^{\rm init}$ at 90 and 60 pc resolution.
\edit{\textbf{Middle:} From left to right, the fraction of current gas and stars in young stars, the moderate correlation (Spearman $\rho = 0.42$) and large scatter between $M_{\rm New}$ and $ \langle \Sigma_{\rm Mol} \rangle_{150 \rm \, pc}$, and three estimates of $M_{\rm New} / M_{\rm Cloud}$ vs $ \langle \Sigma_{\rm Mol} \rangle_{150 \rm \, pc}$}.
\textbf{Bottom:} From left to right, the fraction of initial gas seen in stars and ionized gas, fraction of depleted initial molecular gas in stars and ionized gas, statistical local depletion in molecular gas towards \ion{H}{2} regions at our fiducial 150 pc resolution, and finally the depletion for smaller samples (where available) at 90 pc (purple) and 60 pc (grey) resolution, relative to the dashed black 1:1 line.
}
\label{fig:SFE-est}
\end{figure*}

\begin{figure*}[t]
\centering
\includegraphics[width=1\textwidth]{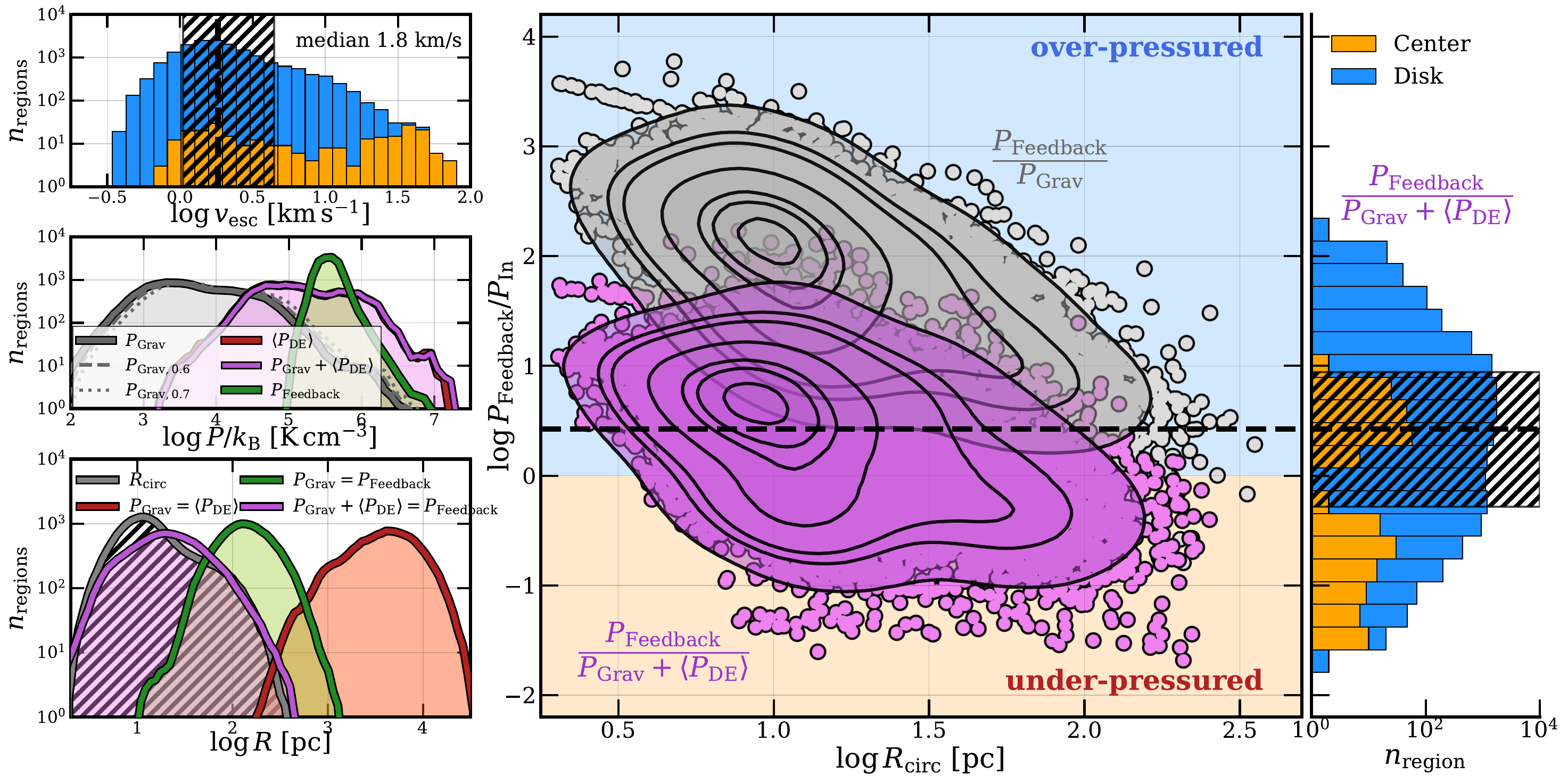}
\caption{\textbf{Top Left:} Distribution of \ion{H}{2} region escape velocities in galaxy centers and disks. \textbf{Middle Left:} Distribution of pressures due to \ion{H}{2} region self-gravity for $M_{\rm sh} = 50\% M_{\rm Mol}^{\rm init}$ (fiducial value, solid grey), $60\% M_{\rm Mol}^{\rm init}$ (dashed grey), and $70\% M_{\rm Mol}^{\rm init}$ (dot-dashed grey); disk $\langle P_{\rm DE} \rangle$ \citep[from][maroon]{2023SUN}, fiducial $P_{\rm Grav}+ \langle P_{\rm DE} \rangle$ (purple), and total $P_{\rm Feedback}$ (green). \textbf{Bottom Left:} \edit{Distribution of HII region $R_{\rm circ}$ (grey hatches) and the sizes at which we expect $P_{\rm Grav} = \langle P_{\rm DE} \rangle$ (maroon), $P_{\rm Grav} = P_{\rm Feedback}$ (green), and $P_{\rm Grav} + \langle P_{\rm DE} \rangle = P_{\rm Feedback}$ (purple).}
\textbf{Right:} \ion{H}{2} region over-pressure (shaded blue) vs under-pressure (shaded orange) for $P_{\rm Feedback}/P_{\rm Grav}$ (grey) and $P_{\rm Feedback}/(P_{\rm Grav} + \langle P_{\rm DE} \rangle)$ (purple, dashed line indicates median, summary histogram to right) with $5^{\rm th}{/}16^{\rm th}{/}25^{\rm th}{/}50^{\rm th}{/}75^{\rm th}{/}84^{\rm th}{/}95^{\rm th}$ percentile contours. 
}
\label{fig:overpressure}
\end{figure*}

\input{intext-table}

\section{Results} \label{sec:discussion}

%\subsection{Comparing Mass Components} 
\subsection{Contribution of \edit{Ionized Gas and Older Stars}} 
\label{sec:comparing-mass-comps}

In Fig.~\ref{fig:Mass-Terms-All4} and Fig.~\ref{fig:Mass-v-R-LHa} we show the masses for our sample, which we summarize in Table~\ref{tab:summary-short} (see Table \ref{tab:summary-all} in Appendix \ref{sec:appendix:summary-tab}). These show that \edit{ionized gas and older stars} contribute significantly to the masses of \ion{H}{2} regions. After taking these into account, the mass of young stars, $M_{\rm New}$, only accounts for $\approx 10-15\%$ of the current mass in a region on average.
\edit{This is in excellent agreement with hydrodynamic simulations of star cluster formation and radiation feedback in turbulent molecular clouds \citep[see e.g.,][]{2018JKIM, 2025MENON}.}

Ionized gas contributes significant mass to \ion{H}{2} regions. \edit{Adopting the Strömgren sphere model, the ionized gas mass} is typically $5\times M_{\rm New}$ (Table \ref{tab:summary-short}). 
% This Strömgren \edit{mass} estimate does not account for clumping within the region, and so represents a lower limit on the density.
\edit{If instead we adopt the value of the electron density implied by the [\ion{S}{2}] doublet, then the ionized gas mass} is $\approx 2\times M_{\rm New}$ (\S \ref{sec:estimating-MHII}). 
%This represents a lower bound on the total ionized gas mass, as discussed in section \S \ref{sec:estimating-MHII}.
% because the [\ion{S}{2}]-derived density likely does not reflect the less dense and more volume-filling ionized gas in the region and biases our statistics towards the densest subset of the Barnes et al. catalog (\S \ref{sec:estimating-MHII}). 
We thus expect that the range $M_{\rm Ion} = 2{-}5 M_{\rm New}$ captures the true value, \edit{once again in agreement with cloud-scale simulations \citep[see][]{2018JKIM, 2021KIM}}. 

\edit{Thus} typical \ion{H}{2} regions contain large reservoirs of gas that has been ionized by pre-supernova feedback, but not yet cleared from the regions. Models of feedback in star-forming regions \citep[e.g.,][]{2019RAHNER,2022GRUDIC,2025LANCASTER} should reproduce this reservoir.

Because \edit{$M_{\rm New} \propto L_{\rm H\alpha}^{\rm corr}$} and $M_{\rm Ion} \propto L_{\rm H\alpha}^{\rm corr} / n_{\rm e}$, denser regions will have lower $M_{\rm Ion}/M_{\rm New}$. Equating Eq.~\eqref{eq:MNew-scaled} and \eqref{eq:estimating-MHII-SII} yields $n_{\rm e} = 81 \rm \, cm^{-3}$ as the density needed to balance $M_{\rm New} = M_{\rm Ion}$. For any region with higher $n_{\rm e}$, $M_{\rm New} > M_{\rm Ion}$. Most regions in PHANGS--MUSE appear to have low $n_{\rm e}$ and hence large ionized gas masses, but density-sensitive spectroscopy targeting high redshift sources suggests much higher $n_{\rm e} \sim 10^2{-}10^3$ cm$^{-3}$ \citep[e.g.,][]{TOPPING2025}. If these spectroscopically inferred densities reflect the volume-averaged values then $M_{\rm Ion}$ will be much less important for the mass budget of these systems.

\ion{H}{2} regions are often considered as ionized nebulae surrounding young stars. \edit{In addition to the large reservoirs of ionized gas, our results indicate that for most $\sim10$~pc \ion{H}{2} regions, $M_{\rm Old} \approx M_{\rm New}$. Because $\rho_{\rm Old} \approx \textrm{constant}$ near each region and $M_{\rm new}$ is fixed by $L_{\rm H\alpha}$, the relative importance of the new and older \textit{stellar} mass} depends on the size scale considered. Fig. \ref{fig:Mass-v-R-LHa} shows that the relative importance of the stellar mass terms changes from $M_{\rm New} \gtrsim M_{\rm Old}$ for small, low--luminosity regions to $M_{\rm Old} > M_{\rm New}$ for larger regions or regions in dense environments like galaxy centers.

\edit{By contrasting $M_{\rm New}$ with the expression for $M_{\rm Old}$ we can identify the scale above which older stars dominate the stellar mass in \ion{H}{2} regions. Equating Eq.~\eqref{eq:MNew-scaled} and \eqref{eq:estimating-MOld-final}, we estimate the size at which we expect $M_{\rm New} = M_{\rm Old}$ for each region, included in the bottom right panel of Fig.~\ref{fig:Mass-v-R-LHa}.} For our sample, the stellar masses balance at median $R = 12 \rm\,pc$ ($16^{\rm th} {-} 84^{\rm th}$ percentile $R \approx 7 {-} 27 \rm\,pc$), and we expect $M_{\rm Old} \gtrsim M_{\rm New}$ on larger scales than this, as previously highlighted in \citet{2023BLACKSTONE}. Thus, \edit{the mass of older stars is already significant on the $\sim 10$~pc scales of \ion{H}{2} regions resolved by HST and may become a more relevant term as} a region is dispersed or evolves into a larger-scale bubble \citep[e.g.,][]{2023EGOROV,2023WATKINS}. \edit{While older stars do not necessarily dominate the overall mass budget on these scales (they have comparable or lower mass than molecular and ionized gas on scales $\lesssim 100$~pc\editt{)}, they represent an important mass component.}

\subsection{Star Formation Efficiencies} \label{sec:SFEs}

In Fig.~\ref{fig:Mass-v-R-LHa}, our estimated initial molecular mass generally exceeds all other components, and $M_{\rm Ion}$ exceeds $M_{\rm New}$. These results indicate the well-known inefficiency of converting gas into stars. The comparison of $M_{\rm New}$ with the initial gas mass before star-formation allows us to estimate the star-formation efficiencies (SFE) $\epsilon$ in each region. $\epsilon$ is defined as the fraction of gas mass converted into stars during the formation of the \ion{H}{2} region \citep[following, e.g.,][]{2007MCKEE}.
% In Fig.~\ref{fig:Mass-v-R-LHa} our estimated initial molecular mass, $M_{\rm Mol}^{\rm init}$, typically exceeds all other components. Likewise $M_{\rm Ion}$ exceeds $M_{\rm New}$. These measurements imply a significant and well-known inefficiency in converting gas to stars. By contrasting $M_{\rm New}$ with estimates of the initial gas mass before star-formation, we can estimate the star-formation efficiencies (SFE) $\epsilon$ for each region. Here $\epsilon$ is defined as the fraction of gas mass that has been converted to stars over the course of forming the \ion{H}{2} region, \citep[following, e.g.,][]{2007MCKEE}, 
\begin{equation}
    \epsilon = \dfrac{M_{\rm New}}{M_{\rm Cloud}^{\rm init}}.
\end{equation}

\paragraph{Upper limit on $\epsilon$ from $M_{\rm Ion}$} 
The initial cloud mass $M_{\rm Cloud}^{\rm init}$ for an individual region can be difficult to constrain, but the current mass of the region $M_{\rm Ion} + M_{\rm New}$  represents a reasonable lower limit under the assumption that no mass has been lost from the region and all of the initial cold gas mass has been converted into stars or photoionized. As shown in Table \ref{tab:summary-short} and Fig. \ref{fig:SFE-est}, contrasting $M_{\rm New}$ and $M_{\rm Ion}+M_{\rm New}$ provides an upper limit on the cluster-scale $\epsilon\approx17\%{-}36\%$ \citep[consistent with, e.g.,][]{2003LADA}.

\paragraph{Lower limit on $\epsilon$ from $M_{\rm Mol}^{\rm init}$} We also contrast $M_{\rm New}$ with our estimated initial molecular gas mass, $M_{\rm Mol}^{\rm init}$. This estimate will include any gas that has been cleared from the region or still remains in the molecular or atomic phase. This calculation yields a minimum $\epsilon\approx6\%$ with $\sim 1$\,dex scatter. This is in line with recent estimates of molecular cloud-scale $\epsilon$ \citep[e.g.,][]{2020CHEVANCE, 2021KIM, 2022KIM, 2023SUN, 2025ZHOU}. 

\edit{Alternatively, folding in the current molecular gas around each region, and setting $M_{\rm Cloud}^{\rm init} = M_{\rm New} + M_{\rm Ion} + M_{\rm Mol}^{\rm current} $ (Fig.~\ref{fig:SFE-est}, middle row, left panel) again yields $\epsilon \approx 6-8\%$, consistent with using the full $M_{\rm Mol}^{\rm init}$.}

\edit{\paragraph{Local Molecular Gas Depletion Towards \ion{H}{2} Regions}}
In Fig. \ref{fig:SFE-est} \edit{(bottom row, right)}, we also compare the current $\Sigma_{\rm mol}$ towards each \ion{H}{2} region to the mass-weighted 1.5 kpc average, $\langle \Sigma_{\rm Mol} \rangle_{\rm 150}$ (which we use to estimate $M_{\rm Mol}^{\rm init}$). \ion{H}{2} regions show lower $\Sigma_{\rm Mol}$ compared to the average $\langle \Sigma_{\rm Mol} \rangle$ in their surroundings, median $\Sigma_{\rm Mol} / \langle \Sigma_{\rm Mol} \rangle_{\rm 150} \approx 0.62$. This statistical depletion of molecular gas towards \ion{H}{2} regions provides another piece of evidence for molecular gas clearing by early stellar feedback (notably, $\approx10\%$ of our \ion{H}{2} regions have no local CO detection at 150\,pc). This allows us to estimate the amount of molecular gas ``missing'' from the region due to either star formation or feedback, $M_{\rm Mol}^{\rm init} - M_{\rm Mol}^{\rm current} \approx 0.38 M_{\rm Mol}^{\rm init}$ with the mass definition following Eq.~\eqref{eq:estimating-MMol}\footnote{Note that the appropriate geometry to translate $\Sigma_{\rm Mol} / \langle \Sigma_{\rm Mol} \rangle_{\rm 150}$ into a missing mass associated with the \ion{H}{2} region is ambiguous. While we use the region size, $\pi R_{\rm circ}^2 (\langle \Sigma_{\rm Mol} \rangle_{\rm 150} - \Sigma_{\rm Mol})$, one could also consider the whole ALMA beam rather than a surface density estimate.}. 

As a test, we repeat the calculation of $\Sigma_{\rm Mol} / \langle \Sigma_{\rm Mol} \rangle$ at increasing physical resolution, from 150 pc (fiducial) to 90 pc and 60 pc (with higher resolution data available only for smaller samples). Fig.~\ref{fig:SFE-est} (bottom right panels) shows that the higher resolution measurements yield qualitatively similar results for local depletion of molecular gas at the locations of \ion{H}{2} regions. $\Sigma_{\rm Mol} / \langle \Sigma_{\rm Mol} \rangle$ near \ion{H}{2} regions steadily drops from 0.62, to 0.60 to 0.53 as the resolution improves from 150 to 90 to 60 pc, respectively. This also implies a statistical constraint on the amount of molecular gas that could be present in a shell around \ion{H}{2} regions. At 60 pc resolution (our best CO$(2-1)$ resolution), the median current molecular gas mass \edit{for our sample is $\lesssim 0.5 M_{\rm Mol}^{\rm init}$. This fraction of molecular gas remaining around \ion{H}{2} regions should evolve over time, and is expected to be larger than our sample median at earlier times.}

\edit{Considering only the depleted molecular gas as the likely initial material for the \ion{H}{2} region, $M_{\rm New}/(M_{\rm Mol}^{\rm init} - M_{\rm Mol}^{\rm current})$, the median $\epsilon$ is $\sim 9\%$ at 150 pc resolution, $7\%$ at 90 pc, and 5\% at 60 pc resolution.}

% Note that the initial molecular gas mass is roughly $20{-}30\times M_{\rm New}$ (Fig.~\ref{fig:Mass-v-R-LHa} and Table \ref{tab:summary-all}). In addition to implying a roughly constant population-averaged SFE on region-scales, more fundamentally, this means $L_{\rm H\alpha}^{\rm corr}/R_{\rm circ}^2 \propto \Sigma_{\rm Mol}$ in \ion{H}{2} regions. 

\paragraph{Photoionization or gas clearing?}
%Despite $M_{\rm New}$ accounting for only a few percent of the initial gas mass, a large amount of the original gas is ionized and still present in \ion{H}{2} regions. 
Contrasting $M_{\rm Ion}$ with $M_{\rm Mol}^{\rm init}$ allows us to estimate the relative importance of photoionization and gas clearing for these regions. The current $M_{\rm New} + M_{\rm Ion}$ accounts for $\approx 33\%$ of the initial gas mass $M_{\rm Mol}^{\rm init} + M_{\rm Atom}^{\rm init}$, as shown in the bottom left panel of Fig.~\ref{fig:SFE-est}. This suggests that on average, $\approx 67\%$ of the original gas has either been dispersed from the region or still resides in the cold phase (perhaps in shells). The other $33\%$ remains mostly as photoionized gas.

We can refine this estimate, as above, by contrasting $M_{\rm New} + M_{\rm Ion}$ with the molecular gas depleted towards \ion{H}{2} regions, $M_{\rm Mol}^{\rm init} - M_{\rm Mol, 150}^{\rm current}$. $M_{\rm New} + M_{\rm Ion}$ accounts for $\approx 55\%$ of this depleted mass. In this case, about half of the molecular gas statistically estimated to be depleted is \edit{still visible as ionized gas and new stars}, while half has been cleared.

% on 1.5\,kpc-scales (used in Eq.~\eqref{eq:estimating-MMol} to estimate $M_{\rm Mol}^{\rm init}$) with 150\,pc-scale measurements at the locations of our \ion{H}{2} regions shows that even on 150\,pc-scales, \ion{H}{2} regions statistically show depletion of molecular gas (see inset panel in Fig.~\ref{fig:SFE-est}). The median local (150\,pc) depletion from the average 1.5\,kpc-scale $\Sigma_{\rm Mol}$ is 53\%.  

% We also measure the current molecular gas surface density near each region and estimate the initial molecular gas mass that was present at the location of the region before star formation. 

\edit{
\paragraph{Variation in $\epsilon$ with $\langle \Sigma_{\rm Mol} \rangle$} From theoretical models, $\epsilon$ is expected to relate to the initial gas surface density $\Sigma_{\rm Cloud}^{\rm init}$ of the molecular cloud core out of which an \ion{H}{2} region is born \citep[e.g.,][and references therein]{2016THOMPSON, 2016RASKUTTI, 2018JKIM, 2021JKIM, 2019HE, 2019GRUDIC, 2020FUKUSHIMA, 2021cLANCASTER, 2022MENON, 2023MENON, 2023CHEVANCE}. Clouds with higher initial surface density are expected to convert more of their material to stars.

%\citet{2025LEROY} and \citet{2025MEIDT} represent recent comprehensive observational constraints on the relationship between $\epsilon$ and molecular gas surface density at $60{-}150$ pc scales for nearby galaxies, finding a $\sim$flat relationship between cloud-scale SFE per free-fall time and $\langle \Sigma_{\rm Mol} \rangle$.

The middle right row in Fig. \ref{fig:SFE-est} shows several $\epsilon$ estimates as a function  $\langle \Sigma_{\rm Mol} \rangle_{150\rm\,pc}$. The figure shows different trends for each approach, reflecting both correlated axes (for $\epsilon$ estimates involving $M_{\rm Mol}^{\rm init}$) and the difficulty of estimating $M^{\rm init}_{\rm Mol}$ for any given region. Still given the relatively low surface densities in our sample, the modest $\epsilon$ observed appears to be in agreement with predictions of efficiencies well below $1$ \citep[e.g.,][]{2016THOMPSON}. Improving estimates of the initial mass and $\epsilon$ represent a goal for future work. For a related exercise focused on the gas depletion time, $M_{\rm Mol}^{\rm init}/{\rm SFR}$, to region-averaged cloud properties see \citet{2025LEROY} or \citet{2025MEIDT}.

%This trend is reproduced in Figure \ref{fig:SFE-est}, where we present the complementary \ion{H}{2} region-scale relationship between $\epsilon$ and $\langle \Sigma_{\rm Mol} \rangle_{150\rm\,pc}$. $M_{\rm New}$ shows significant scatter with $\langle \Sigma_{\rm Mol} \rangle_{\rm 150 \, pc}$, so $M_{\rm New} / M_{\rm Mol}^{\rm init}$ suffers from correlated axes, but yields median values in agreement with e.g., \citet{2016THOMPSON}. 
% At 150 pc, $M_{\rm New}/$ $M_{\rm New}/M_{\rm Mol}^{\rm init}$ declines with $\langle \Sigma_{\rm Mol} \rangle_{\rm 150 \, pc}$ due to correlated axes. 

%However, $M_{\rm New} / (M_{\rm New} + M_{\rm Ion})$ yields interesting results. To first order, the physical density of the ionized gas is correlated with the local molecular gas surface density, and $\epsilon$ agrees with theory given the low surface densities we compute.
%While our $\epsilon$ are in general agreement with predictions from star-formation models, in this Letter we lay the framework for follow up work using much higher resolution (10 pc) CO($2{-}1$) data in Local Group galaxies to better constrain $M_{\rm New} / M_{\rm Cloud}^{\rm init}$ and $\langle \Sigma_{\rm Cloud} \rangle$ in \ion{H}{2} regions at higher resolution.
}

\subsection{Dynamical Evolution of Regions} \label{sec:region-dynamics}

Based on our mass measurements, we provide estimates of the dynamical state of \ion{H}{2} regions. With the enclosed mass $M_{\rm Enclosed} = M_{\rm New} + M_{\rm Ion, Str\ddot{o}mgren} + M_{\rm Old}$ within $R_{\rm circ}$, we first estimate the escape velocity for gas, $v_{\rm esc}$ as,
\begin{equation}
\begin{split} 
v_{\rm esc} &= \biggl ( \dfrac{2GM_{\rm Enclosed}}{R_{\rm circ}} \biggr )^{1/2} \\ 
 &= 2.1 \biggl ( \dfrac{M_{\rm Enclosed} }{5,000 \, M_\odot} \biggr )^{1/2} \biggl ( \dfrac{R_{\rm circ}}{10 \,\rm pc} \biggr )^{-1/2} {\rm km \, s^{-1}}.
\end{split} 
\end{equation}
We calculate a median \ion{H}{2} region $v_{\rm esc}=1.8 \rm\, km \, s^{-1}$, which rises to $\approx 50\rm\, km \, s^{-1}$ for the brightest regions in galaxy centers (Fig.~\ref{fig:overpressure}, see also Table \ref{tab:summary-all}). Comparing to the typical sound speed $c_{\rm s} \approx 11.6  \rm\,km \, s^{-1}$ for $10,000\rm\, K$ warm gas, or the typical RMS warm gas velocity $v_{\rm rms} \approx \sqrt{3}c_{\rm s}\approx20 \rm\,km \, s^{-1}$, on average $v_{\rm esc}<v_{\rm rms}$.

However, $v_{\rm esc}$ alone is not enough to determine whether \ion{H}{2} regions are under-pressured (and hence collapsing), in equilibrium, or over-pressured (expanding). \ion{H}{2} regions evolve in the context of their local galactic environment, and a balance of gravity, the ambient ISM pressure $P_{\rm ext}$, and total feedback pressure determines their evolution. The relevant quantity here is the momentum equation for the shell of swept-up material \citep[e.g.,][]{2010MURRAY, 2017RAHNER, 2019RAHNER, 2025LANCASTER}, %for a thin shell
% \begin{equation*}
% \begin{split} 
% \dfrac{dp_{\rm sh}}{dt} &= -F_{\rm Grav} - F_{\rm Ext} + F_{\rm Feedback}, 
% \end{split} 
% \end{equation*}
which we write here terms of pressure on a thin shell of radius $R_{\rm circ}$,
\begin{equation}
\begin{split} 
\dfrac{1}{4\pi R_{\rm circ}^2}\dfrac{dp_{\rm sh}}{dt} &= -P_{\rm Grav} - P_{\rm ext} + P_{\rm Feedback}, \\
&= - \dfrac{GM_{\rm sh} \big(M_{\rm Enclosed} + \frac{1}{2} M_{\rm sh} \big)}{R_{\rm circ}^2 (4\pi R_{\rm circ}^2)} - \langle P_{\rm DE} \rangle \\ &\quad + \big(P_{\rm Therm} + P_{\rm Rad} + P_{\rm Wind} + P_{\rm X} \big)~.
\end{split} 
\end{equation}
Here $p_{\rm sh}$ is the momentum of a thin \edit{gaseous} shell (thickness $\ll R_{\rm circ}$) of mass $M_{\rm sh}$. We set $M_{\rm sh}$ to $0.5 M_{\rm Mol}^{\rm init}$ since on average $\sim 50\%$ of the initial molecular gas is locally depleted (see \S\ref{sec:SFEs}).\footnote{\edit{On average $\approx 50\%$ of the molecular gas in a beam towards the \ion{H}{2} region appears depleted. We assume the gas around the region itself has been cleared by a similar factor, with the remaining gas in a shell around the region, and so adopt $M_{\rm sh} = 0.5 \times M_{\rm Mol}^{\rm init}$. Given the coarse resolution of the ALMA data we consider this more reliable than $M_{\rm Mol}^{\rm current}$, but note this as a direction where future higher resolutions observations will improve our view.}}
%instead of $M_{\rm Mol}^{\rm current}$ since $M_{\rm Mol}^{\rm init}$ is a more robust measurement averaging over the local 1.5 kpc GMC population, and and does not suffer from flux completeness issues. This also leverages the strong statistical signal of average molecular gas depletion. $M_{\rm Mol}^{\rm current}$ is calculated from aperture photometry on \ion{H}{2} regions that are much smaller than the size of the ALMA beam (e.g., 150 pc) and it is not obvious how much of the CO emission within an aperture should be associated with the shell around an \ion{H}{2} region.}}
% \footnote{\edit{We set $M_{\rm sh}=0.5 M_{\rm Mol}^{\rm init}$ instead of $M_{\rm Mol}^{\rm current}$, since the statistical depletion in $\Sigma_{\rm Mol}$ across different resolutions is a more reliable signal compared to flux retrieval for sub-resolution apertures. This also anchors $M_{\rm sh}$ in our measurement of $M_{\rm Mol}^{\rm init}$ that does not suffer from completeness issues, compared to our resolution-limited measurement of $M_{\rm Mol}^{\rm current}$ for individual regions.}} 
\edit{$P_{\rm Grav}$ is a pressure-like gravity term that accounts for the shell's response to the gravitational force exerted by the mass enclosed by the thin gaseous shell \footnote{\edit{
We take the evolution of the gaseous shell as representative  of the evolution of the \ion{H}{2} region.  %we are only interested in how the gaseous component of the HII region evolves.  
When the shell is static, we take this as an indication that the \ion{H}{2} region is in equilibrium. This occurs when the outward pressure force is balanced by the gravitational force inward and the external compressive ``weight'' of gas and stars in \editt{a \textit{finite} galaxy disk. For any realistically finite disk, the density effectively goes to zero at some large (but finite) distance $L$. For \ion{H}{2} regions in a stellar disk, $ R_{\rm circ} \ll h_z^{\rm exp} \ll L$, and Gauss's Law for gravitation applies as long as $L$ is finite. Hence, $g = - G M_{\rm Enclosed} / R_{\rm circ}^2$ on the shell}.
For a more detailed derivation of $P_{\rm Grav}$, refer to \S 2 and Appendix A in \citet{1988OSTRIKER&MCKEE}.}} as well as the self-gravity of the shell material itself.}
For the external ISM pressure, we use the $1.5\,\rm kpc$-average flux-weighted 150 pc-scale galactic disk dynamical equilibrium pressure $P_{\rm ext} = \langle P_{\rm DE} \rangle$ from the \citet{2022SUN,2023SUN} mega tables.

We contrast these inward pressure terms ($P_{\rm Grav}$ and $\langle P_{\rm DE} \rangle$) against the combined pre-supernova feedback from young stars, $P_{\rm Feedback}$. This includes the warm gas pressure $P_{\rm Therm}$ and the total radiation pressure $P_{\rm Rad}$ from both UV and IR photons computed for this sample of regions in \citet{2025PATHAK}, as well as first-order estimates for the pressure due to stellar winds $P_{\rm Wind}$ \citep[assuming a momentum-conserving wind; see also][]{2021BARNES, 2025LANCASTER} by scaling the luminosity-weighted \textsc{Starburst99} stellar wind outputs averaged over the first 4 Myr of SSP evolution as
\begin{equation}
\begin{split} 
\dfrac{P_{\rm Wind}}{k_{\rm B}} &= \dfrac{\sqrt{2 \dot{M}_{\rm Wind} L_{\rm mech}}}{4\pi R_{\rm circ}^2 k_{\rm B}} \\ &= 7.7\times 10^3 \biggl ( \dfrac{L_{\rm H\alpha}^{\rm corr}}{10^3 L_\odot} \biggr ) \biggl ( \dfrac{R_{\rm circ}}{10\rm\,pc} \biggr )^{-2}  {\rm K \, cm^{-3}}.  \\
% \dfrac{P_{\rm CR}}{k_{\rm B}} &= \dfrac{\eta_{\rm CR} L_{\rm mech}}{4\pi R_{\rm circ} D_{\rm CR} k_{\rm B}} \\ &= 3.0\times 10^3 \biggl ( \dfrac{L_{\rm H\alpha}^{\rm corr}}{10^3 L_\odot} \biggr ) \biggl ( \dfrac{R_{\rm circ}}{10\rm\,pc} \biggr )^{-2}  {\rm K \, cm^{-3}},
\end{split} 
\end{equation}
% assuming a cosmic ray momentum transfer efficiency $\eta_{\rm CR}=0.1$ and a diffusion coefficient $D_{\rm CR}=10^{27} \rm \, cm \, s^{-1}$. 
Finally we include a constant X-ray pressure for 0.2 keV and $n_{\rm e,X} = 5\times10^{-3} \rm \, cm^{-3}$ hot gas \citep[e.g.,][]{2012MINEO} as,
\begin{equation}
\begin{split} 
\dfrac{P_{\rm X}}{k_{\rm B}} &= 2n_{\rm e,X} T_{\rm X}  = 2.3 \times 10^4 {\, \rm K \, cm^{-3}}.
\end{split} 
\end{equation}
\edit{This sets a minimum pressure ($\ll P_{\rm Therm}$) due to diffuse, X-ray emitting hot gas to account for the thermal pressure from shock-heated gas by supernovae and stellar winds.}
While follow-up work will present more rigorous estimates of $P_{\rm Wind}$ and $P_{\rm X}$, this is a reliable first-order estimate that includes the dominant feedback terms in optically bright \ion{H}{2} regions, where $P_{\rm Therm}/k_{\rm B}\approx 5\times 10^5 \rm {\, K \, cm^{-3}}$ primarily drives the total $P_{\rm Feedback}$ \citep[][A. Barnes et al. in preparation]{2014LOPEZ, 2021BARNES, 2025PATHAK}.

In Fig.~\ref{fig:overpressure} we summarize $P_{\rm Grav}$ and $\langle P_{\rm DE}\rangle$ (inward pressure) against $P_{\rm Feedback}$ (outward pressure). The distributions show that generally $\langle P_{\rm DE}\rangle \gg P_{\rm Grav}$, i.e., the ambient ISM pressure is the major term opposing feedback, and gravity plays a subdominant role at this scale. Varying $M_{\rm sh}$ from $0.5 M_{\rm Mol}^{\rm init}$ (fiducial) to 0.6 and $0.7 M_{\rm Mol}^{\rm init}$ increases $P_{\rm Grav}$ by roughly 30\% and 60\%, \edit{but is still not enough for $P_{\rm Grav}$ on the gaseous shell to be significant compared to external $\langle P_{\rm DE}\rangle$. 
%However, since $\langle P_{\rm DE}\rangle\sim\text{constant}$ and $P_{\rm Grav}$ increases with size, we expect $P_{\rm Grav}$ to be a more relevant term for larger-scale bubbles \citep[e.g.,][]{2023WATKINS}. 
In fact, given the local $L_{\rm H\alpha}^{\rm corr}$, $\langle \Sigma_{\rm Mol}\rangle$, and $\rho_{\rm Old}$, the minimum scales at which we expect $P_{\rm Grav} \approx \langle P_{\rm DE} \rangle$  is large, typically a few kpc (see Table \ref{tab:summary-all}).}

\edit{While $P_{\rm Grav} \ll \langle P_{\rm DE} \rangle$, we can still calculate at what scales gravity alone might be comparable to feedback. We expect $P_{\rm Grav} \approx P_{\rm Feedback}$, at $130\rm\,pc$ scales (Fig.~\ref{fig:overpressure}, bottom left). This scale is smaller (median $60\rm\,pc$) for high-density regions in galaxy centers. 
%Because $P_{\rm Grav} \ll \langle P_{\rm DE} \rangle$ on \ion{H}{2} region scales, the overall scale for dynamical balance is smaller. 
By contrast, for most regions, we expect $P_{\rm Grav} + \langle P_{\rm DE} \rangle \approx P_{\rm Feedback}$ at $\approx 20-30$ pc scales, which is $\sim R_{\rm circ}$ (Table \ref{tab:summary-all}).}

We calculate the factor by which regions are ``over-pressured,'' $P_{\rm Feedback}/(P_{\rm Grav} + \langle P_{\rm DE} \rangle)$. Values $>1$ (or log-``over-pressure'' $>0$ in Fig.~\ref{fig:overpressure}) indicate regions that are expanding. The majority ($71\%$) of our \ion{H}{2} regions are over-pressured (expanding), especially smaller regions. Larger regions, especially those in galaxy centers, experience higher $\langle P_{\rm DE} \rangle$ and $P_{\rm Grav}$ and hence become under-pressured despite higher $P_{\rm Feedback}$ \citep[see also][]{2021BARNES}. 

Finally, comparing the ``over-pressure'' with the $R$ at which we expect $P_{\rm Feedback} = P_{\rm Grav} + \langle P_{\rm DE} \rangle$ ($\approx 20\rm\,pc$), and the current $R_{\rm circ}$ of each region confirms that over-pressure/under-pressure corresponds to different stages of \ion{H}{2} region evolution -- early or late -- depending on environment (Table \ref{tab:summary-all}).

% Comparing with the typical sound speed $c_{\rm s} \approx 11.6  \rm\,km \, s^{-1}$ for $10,000\rm\, K$ warm gas, or the typical RMS warm gas velocity $v_{\rm rms} \approx \sqrt{3}c_{\rm s}\approx20 \rm\,km \, s^{-1}$, on average $v_{\rm esc}<v_{\rm rms}$, which implies that most \ion{H}{2} regions are over-pressured and expected to expand at the sound speed $c_{\rm s}$, or lose gas mass via outflows. However, the most luminous \ion{H}{2} regions especially in galaxy centers show $v_{\rm esc}>v_{\rm rms}$, which means that in these regions, the warm gas pressure $P_{\rm Therm}$ alone is likely not sufficient to drive region expansion. These regions are expected to retain the bulk of their ionized gas after cluster formation which may be able to form a second generation of stars, and may contribute to higher cloud-scale SFEs (see \S\ref{sec:SFEs}). 
% Finally, the MUSE $\rm H \alpha$ line widths for our \ion{H}{2} regions are $\approx 30{-}50 \rm\,km \, s^{-1}$, limited by the MUSE line spread function ($\approx 30{-}40 \rm\,km \, s^{-1}$), which means that the MUSE spectral resolution does not allow us to reliably measure the intrinsic velocity dispersion for our \ion{H}{2} regions \citep[][]{2023GROVES}. 

\section{Summary} \label{sec:summary}

Combining high-resolution VLT/MUSE optical spectral mapping, HST H$\alpha$ narrowband imaging, JWST near-IR imaging, ALMA CO data, with VLA and MeerKAT $21\rm\,cm$ data, we measure the masses of $\sim 18,000$ \ion{H}{2} regions across 19 nearby star-forming galaxies at $10\rm\,pc$ physical resolution, including the mass of the young cluster (\S\ref{sec:estimating-MNew}), ionized gas (\S\ref{sec:estimating-MHII}), coincident old stars from the disk (\S\ref{sec:estimating-MOld}), initial molecular gas (\S\ref{sec:estimating-MMol}) and atomic gas (\S\ref{sec:other-mass-comps}), as well as bounds on the dark matter and hot gas mass (\S\ref{sec:other-mass-comps}). These represent the first comprehensive estimates of \ion{H}{2} region masses and self-gravity for a statistically large set of regions outside the Local Group. Our mass estimates have direct implications for the efficiency of stellar feedback in ionizing and clearing cold gas and the dynamical evolution of \ion{H}{2} regions, summarized as follows.  

\begin{enumerate}
    \item The mass of ionized gas in \ion{H}{2} regions is significant, $M_{\rm Ion} = 2{-}5\times M_{\rm New}$, depending on the method used to estimate the ionized gas density (Fig. \ref{fig:Mass-Terms-All4}, \ref{fig:Mass-v-R-LHa}, Table \ref{tab:summary-short}, \S \ref{sec:comparing-mass-comps}). Models of stellar feedback in \ion{H}{2} regions should reproduce significant reservoirs of photoionized but not yet cleared gas.

    \item On the 10\,pc scale of individual regions, the mass of older stars associated with the stellar disk is already comparable to $M_{\rm New}$ (Fig.~\ref{fig:Mass-Terms-All4}, \ref{fig:Mass-v-R-LHa}, Table \ref{tab:summary-short}, \S \ref{sec:comparing-mass-comps}). \edit{Older stars} become \edit{an} increasingly significant \edit{enclosed mass term} at larger scales and in high stellar density environments like galaxy centers. %In the context of region dispersal or the evolution of large-scale shells and bubbles, this older stellar population likely on dominates both the enclosed mass and the region's gravitational potential.
    
    % \item The mass of older stars associated with the stellar disk is already comparable to $M_{\rm New}$ on the 10 pc--scale of individual regions (Fig. \ref{fig:Mass-Terms-All4}, \ref{fig:Mass-v-R-LHa}, Table \ref{tab:summary-short}, \S \ref{sec:comparing-mass-comps}), and increases in importance at larger scales or regions of high stellar density like galaxy centers. When considering dispersal of regions or the evolution of large-scale shells or bubbles, the older stellar component seems certain to dominate the enclosed mass and the gravitational pull of the region.    
    
    \item A conservative upper limit of $\approx 17{-}35\%$ for the \edit{star formation efficiency} (SFE) is obtained by comparing the current mass of young stars and ionized gas, while a lower limit of $\approx 6\%$ results from combining the initial molecular gas and current young stellar mass. The sites of \ion{H}{2} regions show $\approx40{-}50\%$ depletion in molecular gas relative to \edit{the typical surface density in the surrounding region} (\S\ref{sec:SFEs}, Fig~\ref{fig:SFE-est}). 

    \item We compare the effect of gravity to the ISM pressure expected from dynamical equilibrium. When accounting for the full enclosed mass, \edit{self-gravity always plays a secondary role relative to external pressure in confining regions}. Comparing inward and outward pressures, we find that most $R_{\rm circ}\sim10\rm\,pc$ regions are over-pressured relative to their self-gravity and their surroundings and so are likely to expand, while larger regions in galaxy centers appear to be confined (Fig.~\ref{fig:overpressure}, \S \ref{sec:region-dynamics}). Region self-gravity is expected to dominate over feedback pressure at $\approx130\rm\,pc$ scales\edit{, but always remains sub-dominant to the ambient ISM pressure.}

    % \item SFE? 
    % \item dm + Hatom? 
    % \item escape speed?
    
\end{enumerate}

\section*{Acknowledgments} \label{sec:acknowledgements}

\editt{We thank the anonymous referee for their helpful comments.}

This work has been carried out as part of the PHANGS collaboration. This work is based on observations made with the NASA/ESA/CSA JWST.The data were obtained from the Mikulski Archive for Space Telescopes at the Space Telescope Science Institute, which is operated by the Association of Universities for Research in Astronomy, Inc., under NASA contract NAS 5-03127 for JWST. These observations are associated with program 2107. \edit{The specific JWST observations analyzed can be accessed via doi:10.17909/ew88-jt15. The PHANGS-MUSE LP data can be accessed at ESO doi:10.18727/archive/47 and the HST data can be accessed through MAST at doi:10.17909/t9-r08f-dq31. }

This work is also based on observations collected at the European Southern Observatory under ESO programs 094.C-0623 (PI: Kreckel), 095.C-0473,  098.C-0484 (PI: Blanc), 1100.B-0651 (PHANGS-MUSE; PI: Schinnerer), as well as 094.B-0321 (MAGNUM; PI: Marconi), 099.B-0242, 0100.B-0116, 098.B-0551 (MAD; PI: Carollo) and 097.B-0640 (TIMER; PI: Gadotti).

D.P. is supported by the NSF GRFP.

A.K.L. and D.P. gratefully acknowledge support from NSF AST AWD 2205628, JWST-GO-02107.009-A, and JWST-GO-03707.001-A. A.K.L. also gratefully acknowledges support by a Humbolt Research Award. 

L.A.L. acknowledges support through the Heising-Simons Foundation grant 2022-3533.

J.S. acknowledges support by the National Aeronautics and Space Administration (NASA) through the NASA Hubble Fellowship grant HST-HF2-51544 awarded by the Space Telescope Science Institute (STScI), which is operated by the Association of Universities for Research in Astronomy, Inc., under contract NAS~5-26555.  

DJP greatly acknowledges support from the South African Research Chairs Initiative of the Department of Science and Technology and National Research Foundation.

MB acknowledges support from the ANID BASAL project FB210003. This work was supported by the French government through the France 2030 investment plan managed by the National Research Agency (ANR), as part of the Initiative of Excellence of Université Côte d’Azur under reference number ANR-15-IDEX-01.

ZB and DC gratefully acknowledge the Collaborative Research Center 1601 (SFB 1601 sub-project B3) funded by the Deutsche Forschungsgemeinschaft (DFG, German Research Foundation) – 500700252.  DC acknowledges support by the Deut\-sche For\-schungs\-ge\-mein\-schaft, DFG project number SFB956-A3.

KG is supported by the Australian Research Council through the Discovery Early Career Researcher Award (DECRA) Fellowship (project number DE220100766) funded by the Australian Government. 

ER acknowledges the support of the Natural Sciences and Engineering Research Council of Canada (NSERC), funding reference number RGPIN-2022-03499 and support from the Canadian Space Agency, funding reference 23JWGO2A07.

OE acknowledges funding from the Deutsche Forschungsgemeinschaft (DFG, German Research Foundation) -- project-ID 541068876.

PSB acknowledges  support from the Spanish grant PID2022-138855NB-C31,
funded by MCIN/AEI/10.13039/501100011033/FEDER, EU.

RSK acknowledges financial support from the ERC via Synergy Grant ``ECOGAL'' (project ID 855130),  from the German Excellence Strategy via the Heidelberg Cluster ``STRUCTURES'' (EXC 2181 - 390900948), and from the German Ministry for Economic Affairs and Climate Action in project ``MAINN'' (funding ID 50OO2206). 

\facilities{JWST, VLT/MUSE, HST, ALMA, VLA, MeerKAT}

\software{astropy \citep{ASTROPY13,ASTROPY18}}

\bibliography{main}{}
\bibliographystyle{aasjournalv7}

\appendix
\section{Summary of \ion{H}{2} Region Properties} \label{sec:appendix:summary-tab}
\input{summary-table}

\input{catalog-table}

% \section*{Stuff} 
% \input{temp-equation-dump}

% \allauthors

\end{document}

%% file: authors.tex
% Please add your name (and affiliation) here if you would like to join as a co-author! Thank you in advance :)

% -------------------

\author[0000-0003-2721-487X]{Debosmita Pathak}
\affiliation{Department of Astronomy, Ohio State University, 180 W. 18th Ave, Columbus, Ohio 43210}
\affiliation{Center for Cosmology and Astroparticle Physics, 191 West Woodruff Avenue, Columbus, OH 43210, USA}
\email[show]{pathak.89@buckeyemail.osu.edu}

\author[0000-0002-2545-1700]{Adam K. Leroy}
\affiliation{Department of Astronomy, Ohio State University, 180 W. 18th Ave, Columbus, Ohio 43210}
\affiliation{Center for Cosmology and Astroparticle Physics, 191 West Woodruff Avenue, Columbus, OH 43210, USA}
\email[]{leroy.42@osu.edu}

\author[0000-0003-0410-4504]{Ashley.~T.~Barnes}
\affiliation{European Southern Observatory, Karl-Schwarzschild-Stra{\ss}e 2, 85748 Garching, Germany}
\email[]{ashley.barnes@eso.org}

\author[0000-0003-2377-9574]{Todd A. Thompson}
\affiliation{Department of Astronomy, Ohio State University, 180 W. 18th Ave, Columbus, Ohio 43210}
\affiliation{Center for Cosmology and Astroparticle Physics, 191 West Woodruff Avenue, Columbus, OH 43210, USA}
\affiliation{Department of Physics, Ohio State University, 91 West Woodruff Ave
Columbus, Ohio 43210}
\email[]{thompson.1847@osu.edu}

\author[0000-0002-1790-3148]{Laura A. Lopez}
\affiliation{Department of Astronomy, Ohio State University, 180 W. 18th Ave, Columbus, Ohio 43210}
\affiliation{Center for Cosmology and Astroparticle Physics, 191 West Woodruff Avenue, Columbus, OH 43210, USA}
\email[]{lopez.513@osu.edu}

% ----- Please add your name + orcID + affiliation + EMAIL below! ----- %

% !! NOTE: PLEASE REMEMBER TO ADD YOUR EMAIL !! The overleaf won't compile without an email for each author :(

\author[0000-0002-4378-8534]{Karin M. Sandstrom}
\affiliation{Department of Astronomy \& Astrophysics, University of California, San Diego, 9500 Gilman Dr., La Jolla, CA 92093, USA}
\email[]{kmsandstrom@ucsd.edu}

\author[0000-0003-0378-4667]{Jiayi~Sun}
\altaffiliation{NASA Hubble Fellow}
\affiliation{Department of Astrophysical Sciences, Princeton University, 4 Ivy Lane, Princeton, NJ 08544, USA}
\affiliation{Department of Physics and Astronomy, University of Kentucky, 505 Rose Street, Lexington, KY 40506, USA}
\email{jysun.princeton@gmail.com}

\author[0000-0001-6708-1317]{Simon C.~O.\ Glover}
\affiliation{Universit\"{a}t Heidelberg, Zentrum f\"{u}r Astronomie, Institut f\"{u}r Theoretische Astrophysik, Albert-Ueberle-Str.\ 2, 69120 Heidelberg, Germany}\email[]{glover@uni-heidelberg.de}

\author[0000-0002-0560-3172]{Ralf S.\ Klessen}
\affiliation{Universit\"{a}t Heidelberg, Zentrum f\"{u}r Astronomie, Institut f\"{u}r Theoretische Astrophysik, Albert-Ueberle-Str.\ 2, 69120 Heidelberg, Germany}
\affiliation{Universit\"{a}t Heidelberg, Interdisziplin\"{a}res Zentrum f\"{u}r Wissenschaftliches Rechnen, Im Neuenheimer Feld 225, 69120 Heidelberg, Germany}
\email{klessen@uni-heidelberg.de}

\author[0000-0001-9605-780X]{Eric W. Koch}
\affiliation{National Radio Astronomy Observatory, 800 Bradbury SE, Suite 235, Albuquerque, NM 87106 USA}
\affiliation{Center for Astrophysics $\mid$ Harvard \& Smithsonian, 60 Garden St., 02138 Cambridge, MA, USA}
\email[]{koch.eric.w@gmail.com}

\author[0000-0003-3917-6460]{Kirsten~L.~Larson}
\affiliation{AURA for the European Space Agency (ESA), Space Telescope Science Institute, 3700 San Martin Drive, Baltimore, MD 21218, USA}
\email[]{kilarson@stsci.edu}

\author[0000-0002-2278-9407]{Janice Lee}
\affiliation{AURA for the European Space Agency (ESA), Space Telescope Science Institute, 3700 San Martin Drive, Baltimore, MD 21218, USA}
\email[]{jlee@stsci.edu}

\author[0000-0002-6118-4048]{Sharon Meidt}
\affiliation{Sterrenkundig Observatorium, Universiteit Gent, Krijgslaan 281 S9, B-9000 Gent, Belgium}
\email[]{shmeidt@gmail.com}

\author[0000-0003-0651-0098]{Patricia Sanchez-Blazquez}
\affiliation{Facultad de CC Físicas \& IPARCOS, Universidad Complutense de Madrid, Plaza de las Ciencias 1, 28040, Madrid}
\email[]{psanchezblazquez@ucm.es}

\author[0000-0002-3933-7677]{Eva Schinnerer}
\affiliation{Max-Planck-Institut für Astronomie, Königstuhl 17, D-69117 Heidelberg, Germany}
\email[]{schinner@mpia.de}

\author[0009-0001-1221-0975]{Zein Bazzi}\affiliation{Argelander-Institut f\"ur Astronomie, University of Bonn, Auf dem H\"ugel 71, 53121 Bonn, Germany}\email[]{zbazzi@uni-bonn.de}

\author[0000-0002-2545-5752]{Francesco Belfiore}
\affiliation{INAF -- Osservatorio Astrofisico di Arcetri, Largo E. Fermi 5, I-50157, Firenze, Italy}
\email[]{francesco.belfiore@inaf.it}

\author[0000-0003-0946-6176]{Médéric~Boquien}
\affiliation{Université Côte d'Azur, Observatoire de la Côte d'Azur, CNRS, Laboratoire Lagrange, 06000, Nice, France}
\email[]{email@email.com}

\author[0000-0001-8241-7704]{Ryan Chown}
\affiliation{Department of Astronomy, Ohio State University, 180 W. 18th Ave, Columbus, Ohio 43210}
\email[]{chown.5@osu.edu}

\author[0000-0001-9793-6400]{Dario Colombo}
\affiliation{Argelander-Institut f\"ur Astronomie, University of Bonn, Auf dem H\"ugel 71, 53121 Bonn, Germany}
\email[]{dcolombo@uni-bonn.de}

\author[0000-0002-8549-4083]{Enrico Congiu}
\affiliation{European Southern Observatory (ESO), Alonso de Córdova 3107, Casilla 19, Santiago 19001, Chile}
\email[]{econgiu@eso.org}

\author[0000-0002-4755-118X]{Oleg~V.~Egorov}\affiliation{Astronomisches Rechen-Institut, Zentrum f\"{u}r Astronomie der Universit\"{a}t Heidelberg, M\"{o}nchhofstra\ss e 12-14, D-69120 Heidelberg, Germany}\email[]{oleg.egorov@uni-heidelberg.de}

\author[0000-0002-1185-2810]{Cosima Eibensteiner}
\altaffiliation{Jansky Fellow of the National Radio Astronomy Observatory}
\affiliation{National Radio Astronomy Observatory, 520 Edgemont Road, Charlottesville, VA 22903, USA}
\email{ceibenst@nrao.edu, cosimaeibensteiner.astro@gmail.com}

\author[0000-0001-6615-5492]{Sushma Kurapati}
\affiliation{Netherlands Institute for Radio Astronomy (ASTRON), Oude Hoogeveensedijk 4, 7991 PD Dwingeloo, the Netherlands}
\email[]{kurapati@astron.nl}

\author[0000-0002-0472-1011]{Miguel Querejeta}
\affiliation{Observatorio Astron{\'o}mico Nacional (IGN), C/Alfonso XII 3, Madrid E-28014, Spain}
\email[]{m.querejeta@oan.es}

\author[0000-0002-5782-9093]{Daniel~A.~Dale}
\affiliation{Department of Physics and Astronomy, University of Wyoming, Laramie, WY 82071, USA}
\email[]{email@email.com}

\author[0000-0003-0955-9102]{Timo Kravtsov}
\affiliation{Department of Physics and Astronomy, University of Turku, 20014 Turku, Finland}
\email[]{thtkra@utu.fi}

\author[0000-0002-3472-0490]{Mansi Padave}
\affiliation{Department of Astronomy \& Astrophysics, University of California, San Diego, 9500 Gilman Dr., La Jolla, CA 92093, USA}
\email[]{mpadave@ucsd.edu}

\author[0000-0001-7996-7860]{D.J. Pisano}
\affiliation{University of Cape Town, Private Bag X3, Rondebosch, 7701, Republic of South Africa}
\email[]{dj.pisano@uct.ac.za}

\author[0000-0002-5204-2259]{Erik Rosolowsky}
\affiliation{Dept. of Physics, University of Alberta, 4-183 CCIS, Edmonton, Alberta, T6G 2E1, Canada}
\email[]{rosolowsky@ualberta.ca}

\author[0000-0002-6313-4597]{Sumit K. Sarbadhicary}
\affiliation{Department of Physics and Astronomy, The Johns Hopkins University, Baltimore, MD 21218 USA}
\affiliation{Department of Physics, The Ohio State University, Columbus, Ohio 43210, USA}
\affiliation{Center for Cosmology \& Astro-Particle Physics, The Ohio State University, Columbus, Ohio 43210, USA}
\email[]{ssarbad1@jh.edu}

\author[0000-0002-0012-2142]{Thomas~G.~Williams}
\affiliation{Sub-department of Astrophysics, Department of Physics, University of Oxford, Keble Road, Oxford OX1 3RH, UK}
\email[]{thomas.williams@physics.ox.ac.uk}

\author[0000-0002-4663-6827]{Remy Indebetouw}
\affiliation{Astronomy Deptartment, University of Virginia, P.O. Box 400325, Charlottesville, VA, 22904}
\email[]{remy@virginia.edu}

\author[0000-0002-1370-6964]{Hsi-An Pan}
\affiliation{Department of Physics, Tamkang University, No.151, Yingzhuan Road, Tamsui District, New Taipei City 251301, Taiwan}
\email[]{hapan@gms.tku.edu.tw}

\author[0000-0001-7130-2880]{Leonardo \'Ubeda}
\affiliation{Space Telescope Science Insititue, Baltimore, Maryland 21218}
\email[]{lubeda@stsci.edu}

\author[0000-0002-8553-1964]{Amirnezam Amiri}
\affiliation{Department of Physics, University of Arkansas, 226 Physics Building, 825 West Dickson Street, Fayetteville, AR 72701, USA}
\email[]{amirnezamamiri@gmail.com}

\author[0000-0003-0166-9745]{Frank Bigiel}\affiliation{Argelander-Institut f\"ur Astronomie, University of Bonn, Auf dem H\"ugel 71, 53121 Bonn, Germany}\email[]{bigiel@astro.uni-bonn.de}

\author[0000-0003-4218-3944]{Guillermo A. Blanc}
\affiliation{Observatories of the Carnegie Institution for Science, 813 Santa Barbara Street, Pasadena, CA 91101, USA}
\affiliation{Departamento de Astronom\'{i}a, Universidad de Chile, Camino del Observatorio 1515, Las Condes, Santiago, Chile}
\email[]{email@email.com}

\author[0000-0002-3247-5321]{Kathryn~Grasha}
% \altaffiliation{ARC DECRA Fellow}
\affiliation{Research School of Astronomy and Astrophysics, Australian National University, Canberra, ACT 2611, Australia}   
\email[]{kathryn.grasha@anu.edu.au}

% % Example:
% \author[0000-0000-0000-0000]{PHANGS}
% \affiliation{everywhere}
% \email[]{email@email.com}

%% file: intext-table.tex
\begin{deluxetable*}{lclc}[th!]
\tabletypesize{\small}
\tablecaption{Masses, mass ratios, SFEs, and \ion{H}{2} region pressures \label{tab:summary-short}}
\tablewidth{1\textwidth}

\tablehead{
\textsc{Quantity} & \textsc{Formula} & \textsc{Unit} & \textsc{Median} 
% \multicolumn{2}{c}{\textsc{Quantity}} & \multicolumn{4}{c}{\textsc{Unweighted}} & \multicolumn{4}{c}{\textsc{$L_{\rm H\alpha}^{\rm corr}$-weighted}} \\ \hline
}
\startdata  
% mass estimates summarized
\multicolumn{4}{l}{ {\textsc{Masses}} (\S\ref{sec:estimating-mass-comps}) } \\  \hline \\ [-8pt]
Young Stars (\S\ref{sec:estimating-MNew}) & $M_{\rm New} = \dfrac{0.112 M_\odot}{L_\odot} \times \dfrac{L_{\rm H\alpha}^{\rm corr}}{(1-f_{\rm esc})}$ & $\log M_\odot$ & $2.81_{2.12}^{3.88}$ \\ [8pt]
Strömgren Ionized Gas (\S\ref{sec:estimating-MHII}) & $M_{\rm Ion, Str\ddot{o}mgren} = 1.36 \, m_{\rm H} \, n_{\rm H, Str\ddot{o}mgren} \dfrac{4}{3}\pi R_{\rm circ}^3$ & $\log M_\odot$ & $3.5_{2.78}^{4.63}$ \\ [8pt]
$^{\dagger}$[\ion{S}{2}] Ionized Gas (\S\ref{sec:estimating-MHII}) & $ M_{\rm Ion, SII} = 1.36 \, m_{\rm H} \, \dfrac{L_{\rm H\alpha}^{\rm corr}}{0.45 \langle h \nu \rangle_{\rm H\alpha} \alpha_{\rm B} \, n_{\rm e, SII}}$ & $\log M_\odot$ & $4.61_{3.51}^{5.33}$ \\ [8pt]
Older Disk Stars (\S\ref{sec:estimating-MOld}) & $M_{\rm Old} = \dfrac{\Sigma_{\rm Old}}{2h^{\rm exp}_{z}} \dfrac{4}{3}\pi R_{\rm circ}^3 $ & $\log M_\odot$ & $2.8_{1.93}^{4.14}$ \\ [8pt]
Total Enclosed Mass & $M_{\rm Enclosed} = M_{\rm New} + M_{\rm Ion, Str\ddot{o}mgren} + M_{\rm Old}$ & $\log M_\odot$ & $3.67_{2.93}^{4.83}$ \\ [8pt]
Initial Molecular Gas (\S\ref{sec:estimating-MMol}) & $M_{\rm Mol}^{\rm init} = \left< \Sigma_{\rm Mol} \right>_{150} \, \pi R_{\rm circ}^2$ & $\log M_\odot$ & $4.1_{3.33}^{5.12}$ \\ [8pt]
Initial Atomic Gas (\S\ref{sec:other-mass-comps}) & $M_{\rm Atom}^{\rm init} = 1.36 \, m_{\rm HI}\dfrac{N(\rm HI)}{h_{\rm HI}^{\rm FWHM}} \dfrac{4}{3} \pi R_{\rm circ}^3$ & $\log M_\odot$ & $1.83_{0.98}^{3.1}$ \\ [10pt] \hline
% Mass ratios summarized
\multicolumn{4}{l}{{\textsc{Mass Ratios}} (\S\ref{sec:comparing-mass-comps}) } \\  \hline \\ [-8pt]
% \multirow{2}{*}[-2ex]{Young Stars vs Photoionization} & $\dfrac{M_{\rm Ion, Str\ddot{o}mgren}}{M_{\rm New}}$ & ${-}$ & $00^{00}_{00}$ \\ [8pt]
Young Stars vs Strömgren Ionized Gas & $\dfrac{M_{\rm Ion, Str\ddot{o}mgren}}{M_{\rm New}}$ & ${-}$ & $5.01_{3.99}^{6.48}$ \\ [8pt]
$^{\dagger}$Young Stars vs [\ion{S}{2}] Ionized Gas & $\dfrac{M_{\rm Ion, SII}}{M_{\rm New}}$ & ${-}$ & $1.89_{1.0}^{3.13}$ \\ [8pt]
Young Stars vs Older Stars & $\dfrac{M_{\rm Old}}{M_{\rm New}}$ & ${-}$ & $1.01_{0.41}^{3.1}$ \\ [8pt] 
Current vs Initial Mass & $\dfrac{M_{\rm New} + M_{\rm Ion, Str\ddot{o}mgren}}{M_{\rm Mol}^{\rm init} + M_{\rm Atom}^{\rm init}}$ & ${-}$ & $0.33_{0.15}^{0.73}$ \\ [8pt] 
Current Mass vs Molecular Gas Depletion & $\dfrac{M_{\rm New} + M_{\rm Ion, Str\ddot{o}mgren}}{M_{\rm Mol}^{\rm init} - M_{\rm Mol}^{\rm current}}$ & ${-}$ & $0.55_{0.19}^{2.15}$ \\ [8pt] 
$150\rm\,pc$ Molecular Gas Depletion & $\dfrac{M_{\rm Mol}^{\rm init} - M_{\rm Mol}^{\rm current}}{M_{\rm Mol}^{\rm init}}$ & ${-}$ & $0.38_{-0.43}^{0.8}$ \\ [10pt] \hline
% SFEs summarized
\multicolumn{4}{l}{{\textsc{Star-Formation Efficiencies}} (\S\ref{sec:SFEs}) } \\  \hline \\ [-8pt]
% \multirow{2}{*}[-2ex]{Frac. of Current Mass in Stars } & $\epsilon = \dfrac{M_{\rm New}}{M_{\rm New} + M_{\rm Ion, Str\ddot{o}mgren}}$ & ${-}$ & $00^{00}_{00}$ \\ [8pt]
Frac. of Current Enclosed Mass in Stars (Strömgren) & $\epsilon = \dfrac{M_{\rm New}}{M_{\rm New} + M_{\rm Ion, Str\ddot{o}mgren}}$ & ${-}$ & $0.17_{0.13}^{0.2}$ \\ [8pt]
$^{\dagger}$Frac. of Current Enclosed Mass in Stars ([\ion{S}{2}]) & $\epsilon = \dfrac{M_{\rm New}}{M_{\rm New} + M_{\rm Ion, SII}}$ & ${-}$ & $0.35_{0.24}^{0.5}$ \\ [8pt]
\edit{Frac. of Total Current Mass in Stars (Strömgren)} & $\epsilon = \dfrac{M_{\rm New}}{M_{\rm New} + M_{\rm Ion, Str\ddot{o}mgren} + M_{\rm Mol}^{\rm current}}$ & ${-}$ & $0.06_{0.04}^{0.1}$ \\ [8pt]
$^{\dagger}$\edit{Frac. of Total Current Mass in Stars ([\ion{S}{2}])} & $\epsilon = \dfrac{M_{\rm New}}{M_{\rm New} + M_{\rm Ion, SII} + M_{\rm Mol}^{\rm current}}$ & ${-}$ & $0.08_{0.05}^{0.14}$ \\ [8pt]
Frac. of Initial Molecular Mass in Stars & $\epsilon = \dfrac{M_{\rm New}}{M_{\rm Mol}^{\rm init}}$ & ${-}$ & $0.06_{0.03}^{0.13}$ \\ [8pt]
\edit{Frac. of Depleted Molecular Mass in Stars} & $\epsilon = \dfrac{M_{\rm New}}{M_{\rm Mol}^{\rm init} - M_{\rm Mol}^{\rm current}}$ & ${-}$ & $0.09_{0.03}^{0.34}$ \\ [10pt] \hline
% Pressures summarized
\multicolumn{4}{l}{{\textsc{Pressures}} (\S\ref{sec:region-dynamics}) } \\  \hline \\ [-8pt]
Region Self-Gravity & $P_{\rm Grav} = \dfrac{GM_{\rm sh} \big(M_{\rm Enclosed} + \frac{1}{2} M_{\rm sh} \big)}{4\pi R_{\rm circ}^4}$ & $\log \rm K \, cm^{-3}$ & $3.59_{3.03}^{4.39}$ \\ [8pt]
Pre-Supernova Feedback & $P_{\rm Feedback} = P_{\rm Therm} + P_{\rm Rad} + P_{\rm Wind} + P_{\rm X}$ & $\log \rm K \, cm^{-3}$ & $5.49_{5.41}^{5.6}$ \\ [8pt]
Overpressure & $\dfrac{P_{\rm Feedback}}{P_{\rm Grav} + \langle P_{\rm DE} \rangle }$ & ${-}$ & $2.66_{0.52}^{8.8}$ \\ [10pt] \hline \\ [-12.5pt]
\enddata
\tablecomments{
Summarizing key definitions, units, and and main results (as median$_{16^{\rm th} \, \rm percentile}^{84^{\rm th} \, \rm percentile}$) for masses, mass ratios, SFEs, and pressures of 18,000 \ion{H}{2} regions. See Table \ref{tab:summary-all} for more detailed summary, and Table \ref{tab:catalog-columns} for using the full dataset. \\
$^{\dagger}$ From [\ion{S}{2}] doublet $n_{\rm e, SII}$ for a subset of 3,221 \ion{H}{2} regions (\S\ref{sec:estimating-MHII}).
}
\end{deluxetable*}
% \end{rotatetable}

%% file: summary-table.tex
\begin{deluxetable*}{llrrrrl}[th!]
\tabletypesize{\small}
\tablecaption{Summary of \ion{H}{2} region properties \label{tab:summary-all}}
\tablewidth{1\textwidth}

\tablehead{
\multicolumn{2}{c}{\textsc{Quantity}} & \multicolumn{2}{c}{\textsc{Unweighted}} & \multicolumn{2}{c}{\textsc{$L_{\rm H\alpha}^{\rm corr}$-weighted}}  & \textsc{Notes} \\ \hline
  &  unit    & All  & Center & All  & Center & 
}
\startdata  
$R_{\rm circ}$ & pc & $11.92_{6.6}^{30.27}$ & $22.36_{6.63}^{129.47}$ & $124.26_{68.63}^{200.68}$ & $142.47_{98.68}^{266.57}$ & \multirow{2}{13em}{Barnes et al. in review, see also \citet[][]{2023GROVES}}\\
$L_{\rm H\alpha}^{\rm corr}$ & $\log L_{\odot}$ & $3.6_{2.92}^{4.67}$ & $4.32_{2.93}^{6.47}$ & $6.8_{5.7}^{8.0}$ & $7.82_{7.18}^{8.38}$ \\
$n_{\rm e, Str\ddot{o}mgren}$ & $\rm cm^{-3}$ & $15.24_{11.78}^{19.16}$ & $15.83_{10.83}^{20.93}$ & $16.59_{8.68}^{55.56}$ & $30.18_{16.28}^{160.09}$ & \multirow{2}{13em}{\S\ref{sec:estimating-MHII}} \\
$^{\dagger}$$n_{\rm e, SII}$ & $\rm cm^{-3}$ & $42.33_{25.58}^{79.89}$ & $106.45_{55.05}^{228.61}$ & $130.43_{44.22}^{238.85}$ & $185.17_{132.22}^{282.45}$ \\
$I_{\rm F300M}$ & $\rm MJy \, sr^{-1}$ & $0.4_{0.2}^{0.92}$ & $3.52_{1.65}^{9.47}$ & $2.86_{0.49}^{13.23}$ & $8.96_{3.08}^{17.54}$ & \multirow{2}{13em}{\S\ref{sec:estimating-MOld}} \\
$\rho_{\rm Old}$ & $\log M_{\odot} \, \rm pc^{-3}$ & $-1.04_{-1.36}^{-0.65}$ & $-0.02_{-0.32}^{0.37}$ & $-0.65_{-1.05}^{-0.08}$ & $-0.32_{-1.03}^{0.13}$ \\
$\langle \Sigma_{\rm Mol} \rangle_{150}$ & $\log M_{\odot} \, \rm pc^{-2}$ & $1.52_{1.17}^{1.89}$ & $1.52_{1.38}^{2.33}$ & $2.04_{1.69}^{2.72}$ & $2.72_{2.47}^{2.72}$ & \multirow{1}{13em}{\S\ref{sec:estimating-MMol}; \citet[][]{2022SUN}} \\
$\Sigma_{\rm Atom}$ & $\log M_{\odot} \, \rm pc^{-2}$ & $1.01_{0.79}^{1.19}$ & $0.74_{0.65}^{0.95}$ & $1.03_{0.79}^{1.26}$ & $0.8_{0.71}^{0.85}$ & \multirow{2}{13em}{\S\ref{sec:other-mass-comps}} \\
$\rho_{\rm Atom}$ & $\log M_{\odot} \, \rm pc^{-3}$ & $-1.99_{-2.21}^{-1.81}$ & $-2.26_{-2.35}^{-2.05}$ & $-1.97_{-2.21}^{-1.74}$ & $-2.2_{-2.29}^{-2.15}$ \\ [3pt] \hline
$M_{\rm New}$ & $\log M_{\odot}$ & $2.81_{2.12}^{3.88}$ & $3.53_{2.13}^{5.68}$ & $6.0_{4.9}^{7.2}$ & $7.03_{6.38}^{7.59}$ & \S\ref{sec:estimating-MNew}; Fig.~\ref{fig:Mass-Terms-All4} \\
$M_{\rm Ion, Str\ddot{o}mgren}$ & $\log M_{\odot}$ & $3.5_{2.78}^{4.63}$ & $4.3_{2.79}^{6.57}$ & $6.76_{5.71}^{7.29}$ & $7.27_{6.88}^{7.48}$ & \multirow{2}{13em}{\S\ref{sec:estimating-MHII}; Fig.~\ref{fig:Mass-Terms-All4}} \\
$^{\dagger}$$M_{\rm Ion, SII}$ & $\log M_{\odot}$ & $4.61_{3.51}^{5.33}$ & $5.44_{4.86}^{6.01}$ & $5.98_{5.3}^{6.84}$ & $6.49_{5.91}^{7.37}$ \\
$M_{\rm Old}$ & $\log M_{\odot}$ & $2.8_{1.93}^{4.14}$ & $4.61_{2.91}^{7.19}$ & $6.33_{5.32}^{7.35}$ & $6.97_{5.57}^{7.47}$ & \S\ref{sec:estimating-MOld}; Fig.~\ref{fig:Mass-Terms-All4} \\
$M_{\rm Mol}^{\rm init}$ & $\log M_{\odot}$ & $4.1_{3.33}^{5.12}$ & $4.56_{3.43}^{6.71}$ & $6.66_{5.81}^{7.51}$ & $7.51_{7.35}^{8.06}$  & \S\ref{sec:estimating-MMol}; Fig.~\ref{fig:Mass-Terms-All4} \\
$M_{\rm Atom}^{\rm init}$ & $\log M_{\odot}$ & $1.83_{0.98}^{3.1}$ & $2.29_{0.73}^{4.58}$ & $4.78_{3.92}^{5.35}$ & $5.02_{4.48}^{5.23}$  & \S\ref{sec:other-mass-comps}; Fig.~\ref{fig:Mass-Terms-All4} \\ [3pt] \hline \\ [-10pt]
$\dfrac{M_{\rm Ion, Str\ddot{o}mgren}}{M_{\rm New}}$ & ${-}$ & $5.01_{3.99}^{6.48}$ & $4.83_{3.65}^{7.05}$ & $4.6_{1.38}^{8.8}$ & $2.53_{0.48}^{4.69}$  & \multirow{10}{13em}{\S\ref{sec:comparing-mass-comps}; Fig.~\ref{fig:Mass-v-R-LHa}} \\
$^{\dagger}$$\dfrac{M_{\rm Ion, SII}}{M_{\rm New}}$ & ${-}$ & $1.89_{1.0}^{3.13}$ & $0.75_{0.35}^{1.46}$ & $0.61_{0.34}^{1.81}$ & $0.43_{0.28}^{0.61}$ \\
$\dfrac{M_{\rm Old}}{M_{\rm New}}$ & ${-}$ & $1.01_{0.41}^{3.1}$ & $10.35_{3.55}^{52.96}$ & $2.28_{0.24}^{9.78}$ & $1.81_{0.01}^{11.19}$ \\
$\dfrac{M_{\rm Mol}^{\rm init}}{M_{\rm New}}$ & ${-}$ & $17.71_{7.74}^{39.0}$ & $20.17_{10.96}^{32.52}$ & $10.36_{3.07}^{21.0}$ & $9.15_{1.25}^{15.01}$ \\
$^{\dagger}$$\dfrac{M_{\rm Ion, SII}}{M_{\rm Ion, Str\ddot{o}mgren}}$ & ${-}$ & $0.29_{0.17}^{0.47}$ & $0.13_{0.07}^{0.25}$ & $0.2_{0.08}^{0.46}$ & $0.13_{0.07}^{1.27}$ \\
$\dfrac{M_{\rm Ion, Str\ddot{o}mgren}}{M_{\rm Mol}^{\rm init}}$ & ${-}$ & $0.29_{0.13}^{0.66}$ & $0.25_{0.17}^{0.48}$ & $0.55_{0.32}^{0.88}$ & $0.41_{0.32}^{0.85}$ \\
$\log \dfrac{M_{\rm Atom}^{\rm init}}{M_{\rm Mol}^{\rm init}}$ & ${-}$ & $-2.22_{-2.61}^{-1.77}$ & $-2.48_{-2.76}^{-2.1}$ & $-1.7_{-2.17}^{-1.29}$ & $-2.29_{-2.53}^{-2.04}$  \\ [8pt] \hline  \\ [-10pt]
$\dfrac{M_{\rm New}}{M_{\rm New} + M_{\rm \rm Ion, Str\ddot{o}mgren}}$ & ${-}$ & $0.17_{0.13}^{0.2}$ & $0.17_{0.12}^{0.22}$ & $0.18_{0.1}^{0.42}$ & $0.28_{0.18}^{0.68}$ & \multirow{3}{13em}{\S\ref{sec:SFEs}; Fig.~\ref{fig:SFE-est}} \\
$\dfrac{M_{\rm New}}{M_{\rm Mol}^{\rm init}}$ & ${-}$ & $0.06_{0.03}^{0.13}$ & $0.05_{0.03}^{0.09}$ & $0.1_{0.05}^{0.33}$ & $0.11_{0.07}^{0.8}$ \\ [10pt] \hline \\ [-12pt]
$M_{\rm Enclosed}$ & $\log M_{\odot}$ & $3.67_{2.93}^{4.83}$ & $4.81_{3.15}^{7.33}$ & $7.16_{5.96}^{7.76}$ & $7.63_{7.32}^{7.91}$ & \multirow{10}{13em}{\S\ref{sec:region-dynamics}; Fig.~\ref{fig:overpressure}; $\langle P_{\rm DE} \rangle$ from \citet[][]{2022SUN, 2023SUN}} \\
$v_{\rm esc}$ & $\rm km \, s^{-1}$ & $1.83_{1.05}^{4.43}$ & $4.98_{1.4}^{38.06}$ & $28.58_{10.42}^{54.68}$ & $51.01_{37.8}^{70.87}$ \\
$P_{\rm Grav}$ & $\log \rm K \, cm^{-3}$ & $3.59_{3.03}^{4.39}$ & $3.96_{3.43}^{5.8}$ & $5.19_{4.48}^{6.47}$ & $6.41_{6.07}^{6.98}$ \\
$\langle P_{\rm DE} \rangle$ & $\log \rm K \, cm^{-3}$ & $5.07_{4.54}^{5.79}$ & $5.12_{4.85}^{6.22}$ & $5.91_{5.34}^{6.33}$ & $6.79_{6.22}^{6.97}$ \\
$P_{\rm Feedback}$ & $\log \rm K \, cm^{-3}$ & $5.49_{5.41}^{5.6}$ & $5.54_{5.45}^{5.82}$ & $5.93_{5.46}^{6.84}$ & $6.42_{5.96}^{7.63}$ \\
$\log \dfrac{P_{\rm Feedback}}{P_{\rm Grav} + \langle P_{\rm DE} \rangle}$ & ${-}$ & $0.43_{-0.28}^{0.95}$ & $0.35_{-0.8}^{0.66}$ & $-0.26_{-0.71}^{0.26}$ & $-0.97_{-1.5}^{-0.6}$ \\
\edit{$R_{ \langle P_{\rm DE} \rangle = P_{\rm Grav} }$} & pc & $3746.75_{1482.19}^{7503.73}$ & $572.23_{395.55}^{814.02}$ & $514.7_{9.29}^{4169.61}$ & $659.11_{173.42}^{1688.87}$ \\
\edit{$R_{ P_{\rm Feedback} = P_{\rm Grav} }$} & pc & $127.8_{72.46}^{233.8}$ & $61.04_{45.45}^{92.61}$ & $190.43_{91.03}^{290.0}$ & $62.24_{53.69}^{75.88}$ \\
\edit{$R_{ P_{\rm Feedback} = P_{\rm Grav} + \langle P_{\rm DE} \rangle }$} & pc & $19.06_{8.14}^{44.47}$ & $14.39_{8.47}^{29.08}$ & $34.67_{14.13}^{72.12}$ & $13.15_{5.06}^{21.79}$ \\
\edit{$\log \dfrac{ R_{\rm circ} }{ R_{ P_{\rm Feedback} = P_{\rm Grav} + \langle P_{\rm DE} \rangle } }$} & ${-}$ & $-0.21_{-0.57}^{0.3}$ & $-0.17_{-0.39}^{0.7}$ & $0.43_{0.06}^{0.76}$ & $1.14_{0.6}^{1.56}$ \\ [10pt] \hline
\enddata
\tablecomments{
Summary of \ion{H}{2} region properties and key results with references. The unweighted and $L_{\rm H\alpha}^{\rm corr}$-weighted median and ${16^{\rm th}}{-}{84^{\rm th}}$ percentile range (as median$_{16^{\rm th} \, \rm percentile}^{84^{\rm th} \, \rm percentile}$) for each quantity is presented for the full sample and for galaxy centers.\\
$^{\dagger}$ From [\ion{S}{2}] doublet $n_{\rm e, SII}$ for a subset of 3,221 \ion{H}{2} regions (see \S\ref{sec:estimating-MHII} for details).
% circular radius $R_{\rm circ}$, attenuation-corrected $L_{\rm H\alpha}^{\rm corr}$, Str\"{o}mgren electron number density $n_{\rm e, Str\ddot{o}mgren}$ (\S \ref{sec:estimating-MHII}), \ion{S}{2}-emitting electron number density $n_{\rm e, SII}$ (\S \ref{sec:estimating-MHII}), average background F300M intensity (\S \ref{sec:estimating-MOld}), average background old stellar mass density $\rho_{\rm Old}$ (\S \ref{sec:estimating-MOld}), initial mass-weighted ($\Sigma_{\rm \langle Mol \rangle}$) and unweighted ($\Sigma_{\rm Mol }$) molecular mass surface density (\S \ref{sec:estimating-MMol}), initial atomic gas surface density $\Sigma_{\rm Atom}$ and density $\rho_{\rm Atom}$ (\S\ref{sec:other-mass-comps}), young cluster mass $M_{\rm New}$ (\S\ref{sec:estimating-MNew}), Str\"{o}mgren ionized gas mass $M_{\rm HII, Str\ddot{o}mgren}$ and \ion{S}{2}-emitting ionized clump mass $M_{\rm HII, SII}$ (\S\ref{sec:estimating-MHII}), old stellar mass $M_{\rm Old}$ (\S\ref{sec:estimating-MOld}), mass-weighted and unweighted initial molecular mass $M_{\rm \langle Mol \rangle}$ and $M_{\rm Mol}$ (\S\ref{sec:estimating-MMol}), initial atomic gas mass $M_{\rm Atom}$ (\S\ref{sec:other-mass-comps}), ratios between mass components, total enclosed mass $M_{\rm Enclosed}$ and escape velocity $v_{\rm esc}$ (\S\ref{sec:region-dynamics}). 
}
\end{deluxetable*}
% \end{rotatetable}
% [4pt] \hline \\ [-8pt]

%% file: catalog-table.tex
\begin{deluxetable*}{lcl}[th!]
\tabletypesize{\small}
\tablecaption{\edit{Columns in value-added catalog} \label{tab:catalog-columns}}
\tablewidth{1\textwidth}
\tablehead{
\colhead{Column} & \colhead{Unit} & \colhead{Description}
}
\startdata
\textsf{gal\_name                   }  &                       &  Galaxy name \\
\textsf{region\_ID                  }  &                       &  Nebular region ID from \citet{2023GROVES} and A. Barnes et al. (in review) \\
\textsf{environment                 }  &                       &  Local environment following \citet{2021QUEREJETA} \\
\textsf{Rcirc                       }  & pc                    &  $R_{\rm circ}$ following A. Barnes et. al (in review) and \citet[][]{2025PATHAK} \\
\textsf{L\_HA6562\_CORR             }  & $\rm erg \, s^{-1}$   &  $L_{\rm H\alpha}^{\rm corr}$, estimated to match region size \citep[A. Barnes et al. in review,][]{2025PATHAK} \\
\textsf{MNew                        }  &  $M_\odot$            &  $M_{\rm New}$ (\S\ref{sec:estimating-MNew}) \\
\textsf{ne\_Stromgren               }  &  cm$^{-3}$              &  $n_{\rm e, Str\ddot{o}mgren}$ (\S\ref{sec:estimating-MHII}) \\
\textsf{ne\_SII                     }  &  cm$^{-3}$              &  $n_{\rm e, SII}$ (\S\ref{sec:estimating-MHII}) \\
\textsf{MIon\_Stromgren             }  &  $M_\odot$            &  $M_{\rm Ion, Str\ddot{o}mgren}$ (\S\ref{sec:estimating-MHII}) \\
\textsf{MIon\_SII                   }  &  $M_\odot$            &  $M_{\rm Ion, SII}$ (\S\ref{sec:estimating-MHII}) \\
\textsf{rhoOld                      }  &  $M_\odot\rm\,pc^{-3}$ &  $\rho_{\rm Old}$ (\S\ref{sec:estimating-MOld}) \\
\textsf{MOld                        }  &  $M_\odot$            &  $M_{\rm Old}$ (\S\ref{sec:estimating-MOld}) \\
\textsf{$\langle$SigmaMol\_hex\_150pc$\rangle$}  &  $M_\odot\rm\,pc^{-2}$ &  $\langle \Sigma_{\rm Mol} \rangle_{\rm 150 \rm\, pc}$ from \citet[][]{2022SUN} (\S\ref{sec:estimating-MMol}) \\
\textsf{MMol\_init\_150pc           }  &  $M_\odot$            &  $M_{\rm Mol}^{\rm init}$ (\S\ref{sec:estimating-MMol}) \\
\textsf{MMol\_current\_150pc        }  &  $M_\odot$            &  $M_{\rm Mol}^{\rm current}$ (\S\ref{sec:estimating-MMol}) \\
\textsf{MAtom\_init\_2kpc           }  &  $M_\odot$            &  $M_{\rm Atom}^{\rm init}$ (\S\ref{sec:other-mass-comps}) \\
\textsf{PGrav\_kB                   }  &  K cm$^{-3}$          &  $P_{\rm Grav} / k_{\rm B}$ (\S\ref{sec:region-dynamics}) \\
\textsf{PFeedback\_kB               }  &  K cm$^{-3}$          &  $P_{\rm Feedback} / k_{\rm B}$ (\S\ref{sec:region-dynamics}) \\
\textsf{$\langle$PDE\_kB$\rangle$   }  &  K cm$^{-3}$          &  $\langle P_{\rm DE} \rangle / k_{\rm B}$ from \citet[][]{2022SUN, 2023SUN} (\S\ref{sec:region-dynamics}) \\[4pt] \hline
\enddata
\tablecomments{This table complements the nebular catalogs of \citet{2023GROVES} and A. Barnes et al. (in review) with the basic \ion{H}{2} region properties, masses, SFEs, and pressures needed to reproduce most figures. These catalogs can be joined using the \textsf{region\_ID} from \citet{2023GROVES}. We request that users of these catalogs also cite the original measurements from \citet{2023GROVES} and A. Barnes et al. (in review).}
\end{deluxetable*}
% \end{rotatetable}
% [4pt] \hline